\documentclass[]{jfm}

\usepackage{graphicx}
\usepackage{epstopdf,epsfig}
\usepackage{newtxtext}
\usepackage{newtxmath}
\usepackage{natbib}
\usepackage{hyperref}
\hypersetup{
    colorlinks = true,
    urlcolor   = blue,
    citecolor  = black,
}

\newcommand{\RomanNumeralCaps}[1]
\linenumbers

\usepackage{latexsym}
\usepackage{amsmath}
\usepackage{amssymb}
\usepackage{tabularx}
\usepackage{multirow}
\usepackage{subfigure}
\usepackage{calligra}
\usepackage{calrsfs}
\usepackage{mathrsfs}
\usepackage{extarrows}
\usepackage{booktabs}
\usepackage{bm}
\usepackage{color}
\usepackage{xcolor}

\title{Physical significance of artificial numerical noise in direct numerical simulation of turbulence}

\author{
Shijun Liao\aff{1,2} \corresp{\email{sjliao@sjtu.edu.cn}} \footnotemark[2]
\and Shijie Qin\aff{2} \footnotemark[2]
}

\affiliation{
\aff{1} State Key Laboratory of Ocean Engineering, Shanghai 200240, China
\aff{2} School of Ocean and Civil Engineering, Shanghai Jiao Tong University, Shanghai 200240, China
}

\begin{document}
\begin{sloppypar}
\maketitle 

\begin{abstract}
Using clean numerical simulation (CNS) in which artificial numerical noise is negligible over a finite, sufficiently long interval of time, we provide evidence, for the first time, that artificial numerical noise in direct numerical simulation (DNS) of turbulence  is approximately equivalent to thermal fluctuation and/or stochastic environmental noise. This confers physical significance on the artificial numerical noise of DNS of the Navier-Stokes equations. As a result, DNS on a fine mesh should correspond to turbulence under small internal/external physical disturbance, whereas DNS on a sparse mesh corresponds to  turbulent flow under large physical disturbance, respectively. The key point is that: all of them have physical meanings and so are correct in terms of their deterministic physics, even if their statistics are quite different. This is illustrated herein. Our paper provides a positive viewpoint regarding the presence of artificial numerical noise in DNS.
\end{abstract}

\hspace{-0.4cm}{\bf Keyword} numerical noise, thermal fluctuation, turbulence, Kolmogorov flow 

\footnotetext[2]{These authors contributed equally to this work.}

\section{Introduction}

Turbulence is one of the most challenging problems in fluid mechanics. It is widely accepted by the turbulence community that turbulent flows can be well described by the Navier-Stokes (NS) equations, which are related to the fourth millennium problem \citep{MillenniumProblem}. Direct numerical simulation (DNS) \citep{Orszag1970}, which numerically solves the NS equations without any turbulent models, proved to be a milestone in fluid mechanics because it opened the era of numerical experiments \citep{ShePRSA1991, MoinARFM1998, Scardovelli1999ARFM, Coleman2010DNS}. However, artificial numerical noise caused by truncation and round-off errors is unavoidable for all numerical algorithms, including DNS. It was believed that tiny artificial numerical noise in DNS would not grow to reach large scale because of fluid viscosity. 
However, some scientists pointed out that turbulence governed by NS equations should be chaotic \citep{Deissler1986PoF, aurell1996growth, berera2018chaotic}, and thus there exists the initial exponential growth of average error/uncertainty energy for two-dimensional (2D) and three-dimensional (3D) Navier-Stokes turbulence \citep{boffetta2001predictability, boffetta2017chaos, Vassilicos2023JFM}.
Notably, \cite{Qin2022JFM} demonstrated that artificial numerical noise can lead to huge deviations in the DNS of 2D turbulent Rayleigh-B\'{e}nard convection {\em not only} in spatiotemporal trajectory {\em but also} in statistics by considering the much more accurate results given by `clean numerical simulation' (CNS) \citep{Liao2009, Hu2020JCP, Liao2023book}. Unlike DNS that uses double-precision in general, CNS adopts multiple-precision \citep{oyanarte1990mp} with sufficient significant digits to decrease greatly the round-off error, as well as sufficiently high-order Taylor expansion and small enough spacing for the pseudo-spectral method to decrease greatly the truncation errors in time and space, respectively, so that the artificial numerical noise can be limited to a prescribed small level. As a result, artificial numerical noise in CNS can be negligible over a finite but long enough time interval suitable for calculating statistics. Hence, the CNS results lie close to the true solution of the turbulent flow under consideration and can be used as benchmark data. Therefore, one can carry out clean numerical experiments for turbulence using CNS, where the word `clean' implies that the artificial numerical noise is much smaller than the true solution in a finite, prescribed long time interval $[0,T_{c}]$, and thus can be neglected.   Note that the artificial numerical noise might reach a macro-level once the prescribed time $T_{c}$ has been exceeded, thereafter results given by CNS are also badly polluted by artificial numerical noise, too: this is the reason why, unlike DNS,  simulation of CNS is stopped at a finite prescribed time $T_c$.   This is an obvious difference between DNS and CNS.  In fact, from the viewpoint of CNS, a traditional DNS result using double precision often has a very small value of $T_c$ so that DNS is only a special case of CNS  although unfortunately such a small $T_c$  is useless for calculating statistics.

To confirm the above-mentioned findings of \cite{Qin2022JFM} concerning 2D turbulent Rayleigh-B\'{e}nard convection, \cite{Qin2024JOES} recently used CNS to model a 2D Kolmogorov turbulent flow subject to a periodic boundary condition and a periodic initial condition with a kind of spatial symmetry.  In mathematics, it is obvious that the true solution of the corresponding  2D Kolmogorov turbulent flow should have the same spatial symmetry for {\em all} $t > 0$ as the initial condition.  
It was found that the corresponding CNS result indeed maintains the same spatial symmetry as the initial condition throughout the {\em whole} time interval of simulation, clearly indicating that its numerical noise is indeed negligible so that it is indeed a ``clean'' simulation.  However, the DNS result maintains the same spatial symmetry only at the beginning but quickly loses spatial symmetry completely: this clearly indicates that the small-scale artificial numerical noise of DNS, which is random and without spatial symmetry, indeed quickly grows to become large-scale.

Recently, using CNS to solve a kind of 2D turbulent Kolmogorov flow subject to a specially chosen initial condition that contains micro-level disturbances at different orders of magnitude, say, $O\left(10^{-20}\right)$ and $O\left(10^{-40}\right)$, \cite{Liao2024NEC-arXiv, Liao2025JFM} discovered an interesting phenomenon, which they called ``the noise-expansion cascade'' whereby all micro-level disturbances at different orders of magnitude evolute and grow continuously, step by step, as an inverse cascade, to reach the macro-level. It was found that each disturbance could greatly change the characteristics of the 2D turbulent Kolmogorov flow. This clearly indicates that each disturbance must be considered in the NS equations, even if the disturbance is many orders of magnitude smaller than others.

Note that, just like artificial numerical noise, both {\em internal} thermal fluctuation and {\em external} environmental noise are unavoidable in practice. 
The following fundamental questions arise about the relationships between artificial numerical noise and thermal fluctuations \& environmental noise:
\begin{enumerate} 
\item[(A)] Is artificial numerical noise in DNS approximately equivalent to thermal fluctuation and/or stochastic environmental noise?   

\item[(B)] What is the physical significance of artificial numerical noise in DNS? 

\item[(C)] Are there some turbulent flows whose statistics are sensitive to artificial numerical noise? 
\end{enumerate}
To the best of our knowledge, these are presently open questions. In this paper, we will answer them by first using 
CNS to carry out the `clean' numerical experiment for a 2D turbulent Kolmogorov flow, and then comparing the CNS benchmark solution, whose artificial numerical noise is negligible, with DNS predictions where artificial numerical noise quickly grows to the macro-level.                                        

\section{Mathematical model and numerical algorithm}

Since DNS is mostly used to solve Navier-Stokes equations, let us consider here an incompressible flow in a 2D square domain $[0, L]^2$, called Kolmogorov flow \citep{arnold1960seminar, obukhov1983kolmogorov} with the `Kolmogorov forcing' that is stationary, monochromatic and sinusoidally varying in space with forcing scale $n_K$ and amplitude $\chi$.    
Using the length scale $L/2\pi$ and the time scale $\sqrt{L/2\pi\chi}$, the corresponding  non-dimensional Navier-Stokes equations in the form of stream-function $\psi$ read \citep{chandler2013invariant}: 
\begin{align}
& \frac{\partial}{\partial t}\nabla^{2}\psi+\frac{\partial(\psi,\nabla^{2}\psi)}{\partial(x,y)}-\frac{1}{Re}\nabla^{4}\psi+n_K\cos(n_Ky)=0,       \label{eq_psi}
\end{align}
where $t$ denotes the time, $x$ and $y$ are the Cartesian coordinates within $x,y\in[0,2\pi]$,   $\nabla^{4}=\nabla^{2}\nabla^{2}$ where   
$\nabla^{2}$ is the Laplace operator, 
\[ 
 \frac{\partial(a,b)}{\partial(x,y)}=\frac{\partial a}{\partial x}\frac{\partial b}{\partial y}-\frac{\partial b}{\partial x}\frac{\partial a}{\partial y}      
\]
is the Jacobi operator, and  $Re=\frac{\sqrt{\chi}}{\nu}\left(\frac{L}{2\pi}\right)^{{3}/{2}}$ is the Reynolds number, 
in which $\nu$ denotes the fluid kinematic viscosity. We use here the periodic boundary condition
\begin{align}
& \psi(x,y,t)=\psi(x+2\pi,y,t)=\psi(x,y+2\pi,t),       \label{boundary_condition}
\end{align}
the initial condition
\begin{align}
& \psi(x,y,0)=-\frac{1}{2}[\cos(x+y)+\cos(x-y)],       \label{initial_condition}
\end{align}
and the physical parameters $n_K=16$ and $Re=2000$. All of these are exactly the {\em same} as those used by  \cite{Liao2024NEC-arXiv, Liao2025JFM}.

Since DNS is mostly used to solve Navier-Stokes (NS) equations, let us consider here a simple model about the influence of  thermal fluctuation and/or  stochastic environmental noise on turbulence in the frame of NS equations.  According to the theory of statistical physics concerning thermal fluctuation \citep{Landau1959}, the mean square  of velocity fluctuation is given by 
\begin{align}
& \langle u_{th}^2 \rangle = \frac{k_B\langle T \rangle}{V\langle \rho \rangle},    \label{fluctuations}
\end{align}
where the subscript `th' stands for thermal fluctuation, $k_B$ is the Boltzmann constant, $\langle T \rangle$ and $\langle \rho \rangle$ are the mean temperature and mass density.  
In this paper the fluid is water at room temperature (20\,$^{\circ}$C), thus $k_B=1.38\times10^{-23}$\,J/K, $\langle T \rangle = 293.15$\,K, and $\langle \rho \rangle = 998$\,kg/m$^3$.
Besides, $V$ is regarded as a unit volume. For simplicity, the tiny velocity fluctuation $u_{th}$  is regarded here as a kind of Gaussian white noise with standard deviation $\sigma=10^{-10}$.  
For the 2D turbulent flow,  we have 
\begin{equation}
u=-\frac{\partial \psi}{\partial y},  \;\;\; v = \frac{\partial \psi}{\partial x},
\end{equation}
where the stream-function $\psi$ is governed by Eqs.~(\ref{eq_psi}) to (\ref{initial_condition}).  
Therefore, considering the additional velocity fluctuation $u_{th}$ governed by (\ref{fluctuations}), the stream-function $\psi^*$ with thermal fluctuation term is given by 
\begin{equation}
\psi^*(x,y,t) = \int_0^y -(u+u_{th})\,dy = \psi(x,y,t) - \int_0^y u_{th}(x,y,t)\,dy.    \label{psi-fluctuations}
\end{equation}
Here the term $\int_0^y u_{th}\,dy$ is calculated by the It\^{o} stochastic integral \citep{Pavliotis2014}.  
Note that the velocity fluctuation in $y$ direction, i.e. $v_{th}$,  is given by $-\partial (\int_0^y u_{th}(x,y,t)\,dy)/\partial x $.  
Then, $\psi^*(x,y,t)$  is submitted in the governing equation  (\ref{eq_psi}) to calculate $\psi(x,y,t+\Delta t)$ that is further used to gain $\psi^*(x,y,t+\Delta t)$ by (\ref{psi-fluctuations}), and so on. 
Note that one can also regard the term $ - \int_0^y u_{th}(x,y,t)\,dy$ as stochastic environmental noise.   
In this way, both of the  thermal fluctuation of velocity field and/or stochastic environmental noise  could have influence on the turbulent flow due to the chaotic property of turbulence. Note that the law of mass conservation is always satisfied under the  thermal fluctuation of velocity field and/or stochastic environmental  noise, since  the stream-function is used here.  

On the one hand, we solve Eqs.~(\ref{eq_psi}) to (\ref{initial_condition}) {\em plus} random thermal fluctuation and/or  stochastic environmental noise via (\ref{psi-fluctuations}) throughout the whole interval of time $t \in [0, 300]$ by means of CNS, whose artificial numerical noise is negligible. Here the CNS algorithm is based on the pseudo-spectral method with uniform mesh $1024 \times 1024$ in space, the 140th-order Taylor expansion (i.e. $M=140$) with time-step $\Delta t=10^{-3}$ for temporal evolution, and especially 260 significant digits in multiple precision (i.e. $N_s = 260$) for all variables and parameters.    The results are given the name CNS$^*$, where $^*$ denotes that the CNS result is modified by (\ref{psi-fluctuations}) for thermal fluctuation and/or  stochastic environmental noise whose   evolution is governed by the NS equations (\ref{eq_psi}) to (\ref{initial_condition}).  The corresponding CNS algorithm is exactly the {\em same} as that described by \cite{Qin2024JOES} and \cite{Liao2024NEC-arXiv, Liao2025JFM}, and thus is neglected here.     

On the other hand, we solve the same equations (\ref{eq_psi}) to (\ref{initial_condition}) {\em without} thermal fluctuation and/or stochastic environmental noise by means of DNS over the same interval of time $t \in [0, 300]$, during which the artificial numerical noise quickly enlarges to macro-level. Here we adopt DNS using the pseudo-spectral method with the {\em same} uniform mesh $1024 \times 1024$ in space, but the 4th-order Runge-Kutta's method with time-step $\Delta t=10^{-4}$ in temporal evolution, and double precision (i.e. $N_s = 16$ ) for all variables and parameters, whose results are titled DNS in this paper. As illustrated by  \cite{Liao2024NEC-arXiv, Liao2025JFM}, the artificial numerical noise of DNS quickly enlarges to the same order of magnitude as the true solution, in other words, its spatial-temporal trajectory rapidly becomes badly polluted by artificial numerical noise.         

The relationships between artificial numerical noise in DNS and thermal fluctuation \& environmental noise can be investigated in detail by comparing the CNS$^{*}$ result, whose artificial numerical noise is negligible throughout the whole interval of time $t\in[0,300]$, with the DNS result, whose artificial numerical noise rapidly enlarges to the macro-level. Our detailed findings are given below.

\section{Physical essence of artificial numerical noise of DNS}
 
\subsection{Is numerical noise equivalent to thermal fluctuation and/or environmental noise?}

Here, using the above-mentioned 2D turbulent Kolmogorov flow as an example and by means of CNS plus the tiny modification (\ref{psi-fluctuations}) at each time step, we provide evidence that artificial numerical noise of DNS is approximately equivalent to thermal fluctuation and/or stochastic environmental noise. 
       
First of all, it should be emphasized that, if thermal fluctuation is {\em not} considered, the CNS result retains the spatial {\em symmetry} in the whole time interval $t\in[0,300]$, indicating that the numerical noise of the CNS result is indeed negligible throughout  the whole time interval so that it is indeed  ``clean'', as described by \citet{Liao2024NEC-arXiv, Liao2025JFM}.  By contrast, the DNS result quickly loses this spatial symmetry, clearly indicating that it is badly polluted quickly by numerical noise  \citep{Liao2024NEC-arXiv, Liao2025JFM}.  Based on this known fact, we are quite sure that the numerical noise of the CNS result (mentioned below) with thermal fluctuation  is also ``clean'' and reliable in the whole time interval $t\in[0,300]$.

\begin{figure}
    \begin{center}
        \begin{tabular}{cc}
             \subfigure[]{\includegraphics[width=2.0in]{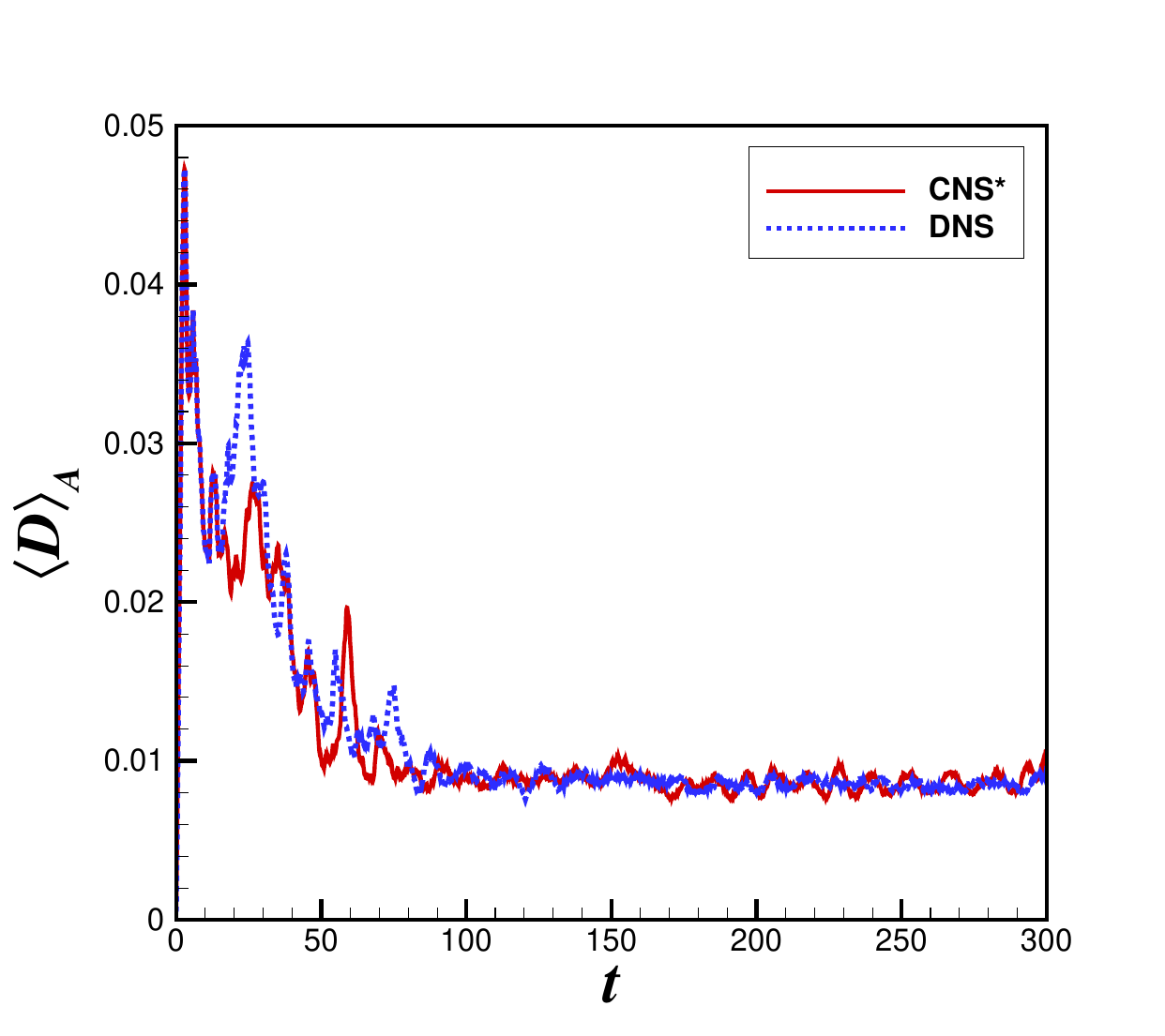}}
             \subfigure[]{\includegraphics[width=2.0in]{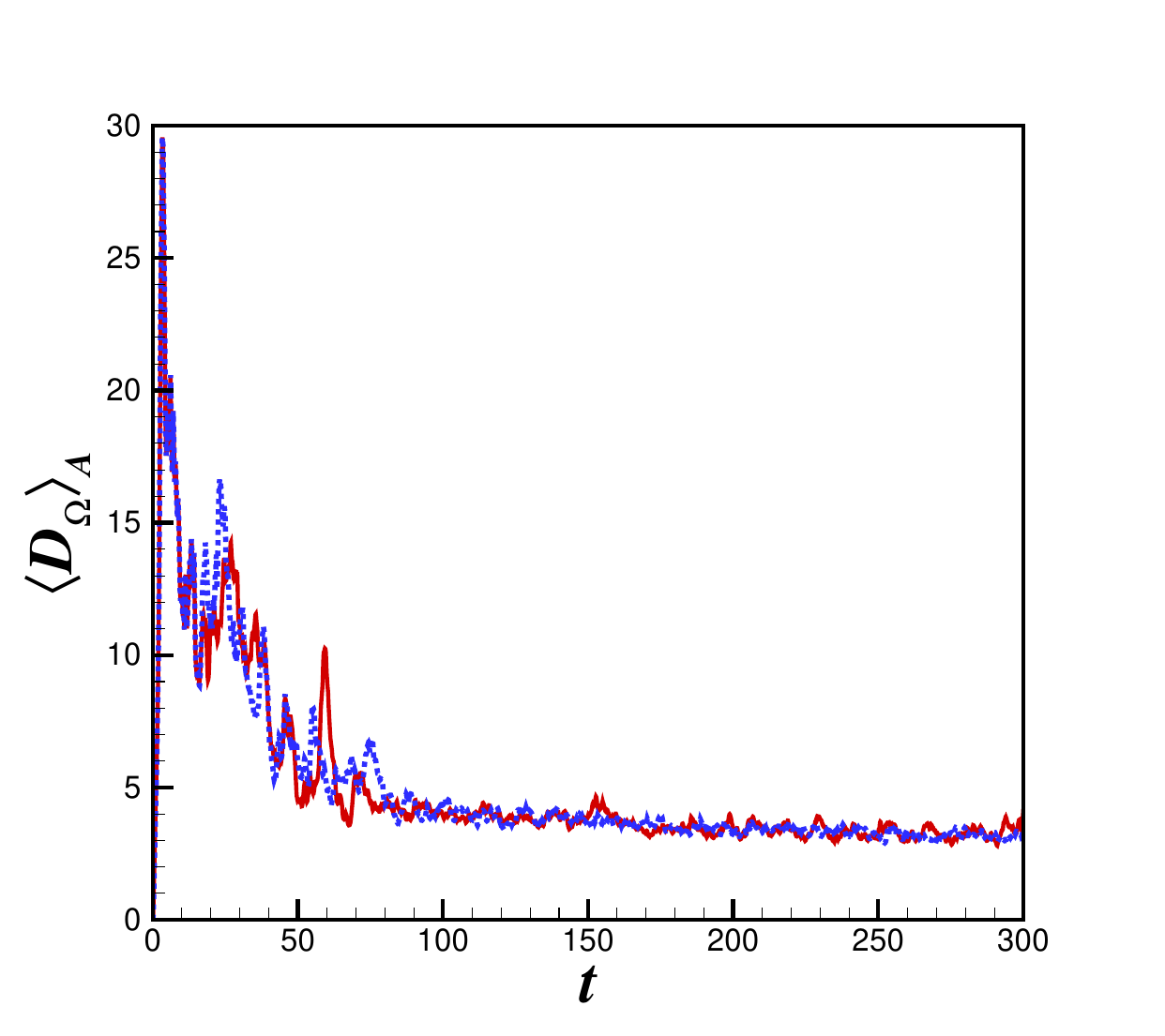}}
        \end{tabular}
    \caption{Time histories of the spatially averaged (a) kinetic energy dissipation rate $\langle D\rangle_A$ and (b) enstrophy dissipation rate $\langle D_{\Omega}\rangle_A$ of the 2D turbulent  Kolmogorov flow: \\CNS$^*$ (red solid line) and DNS (blue dashed line).}     \label{D_t}
    \end{center}
\end{figure}

\begin{figure}
    \begin{center}
        \begin{tabular}{cc}
             \subfigure[]{\includegraphics[width=2.0in]{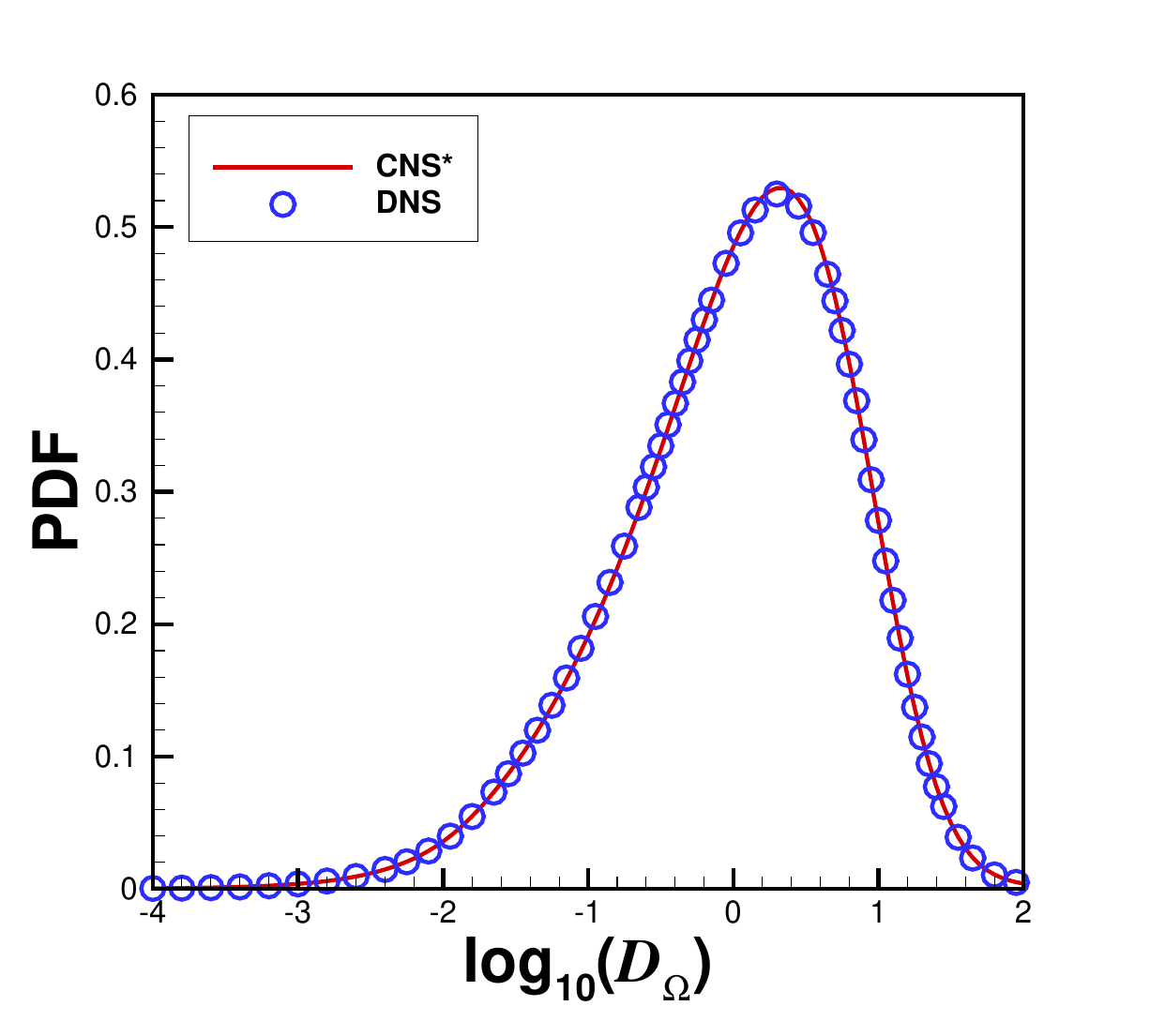}}
             \subfigure[]{\includegraphics[width=2.0in]{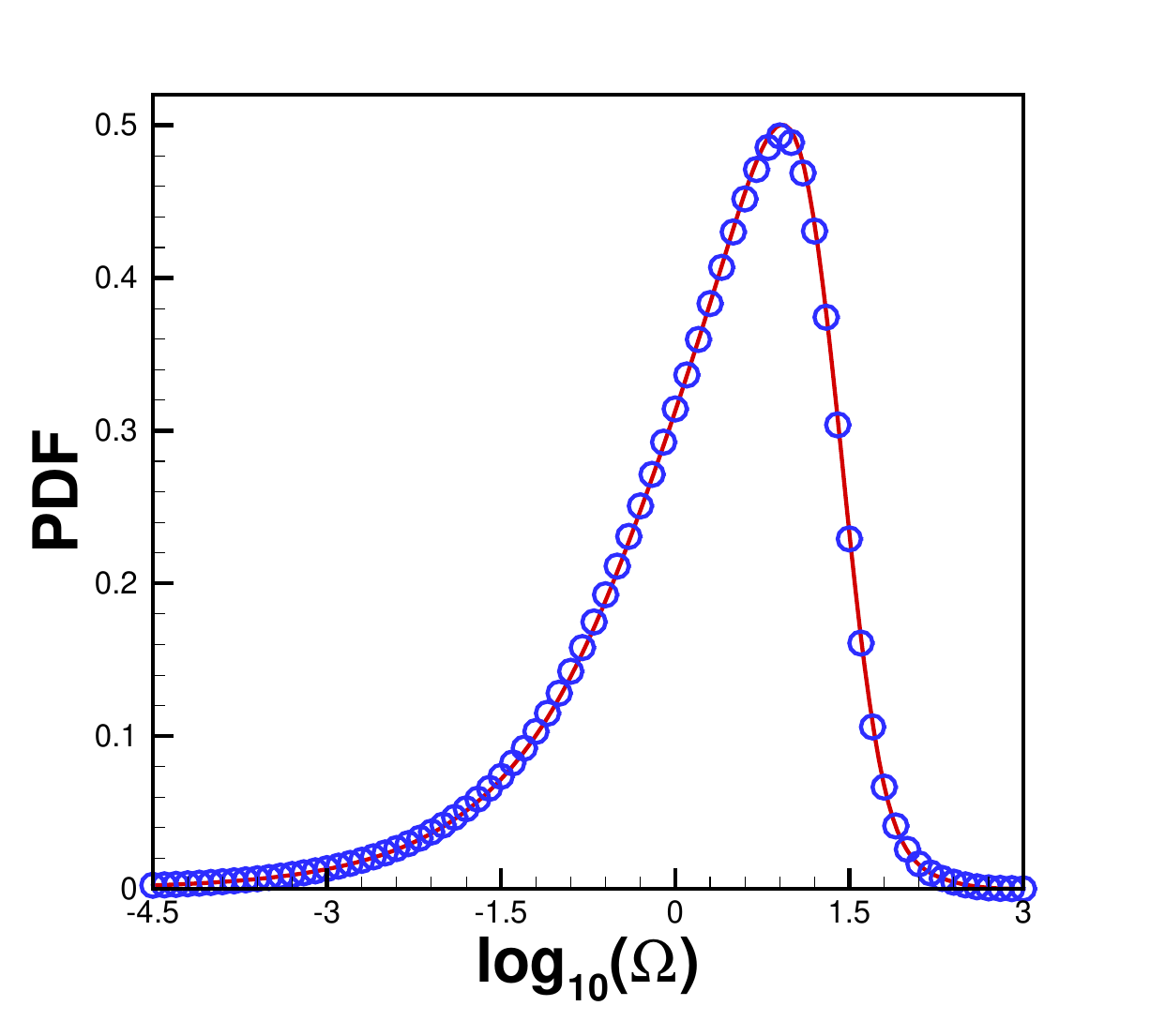}}
        \end{tabular}
    \caption{Probability density functions (PDFs) of (a) the enstrophy dissipation rate $D_\Omega(x,y,t)$ and (b) the enstrophy $\Omega(x,y,t)$ of the 2D turbulent Kolmogorov flow, \\where the integration is taken in $(x,y)\in[0,2\pi)^2$ and $t \in [100, 300]$: \\CNS$^*$ (red line) and DNS (blue circle).}     \label{DEo-PDF}
    \end{center}
\end{figure}

\begin{figure}
    \begin{center}
        \begin{tabular}{cc}
             \subfigure[]{\includegraphics[width=2.0in]{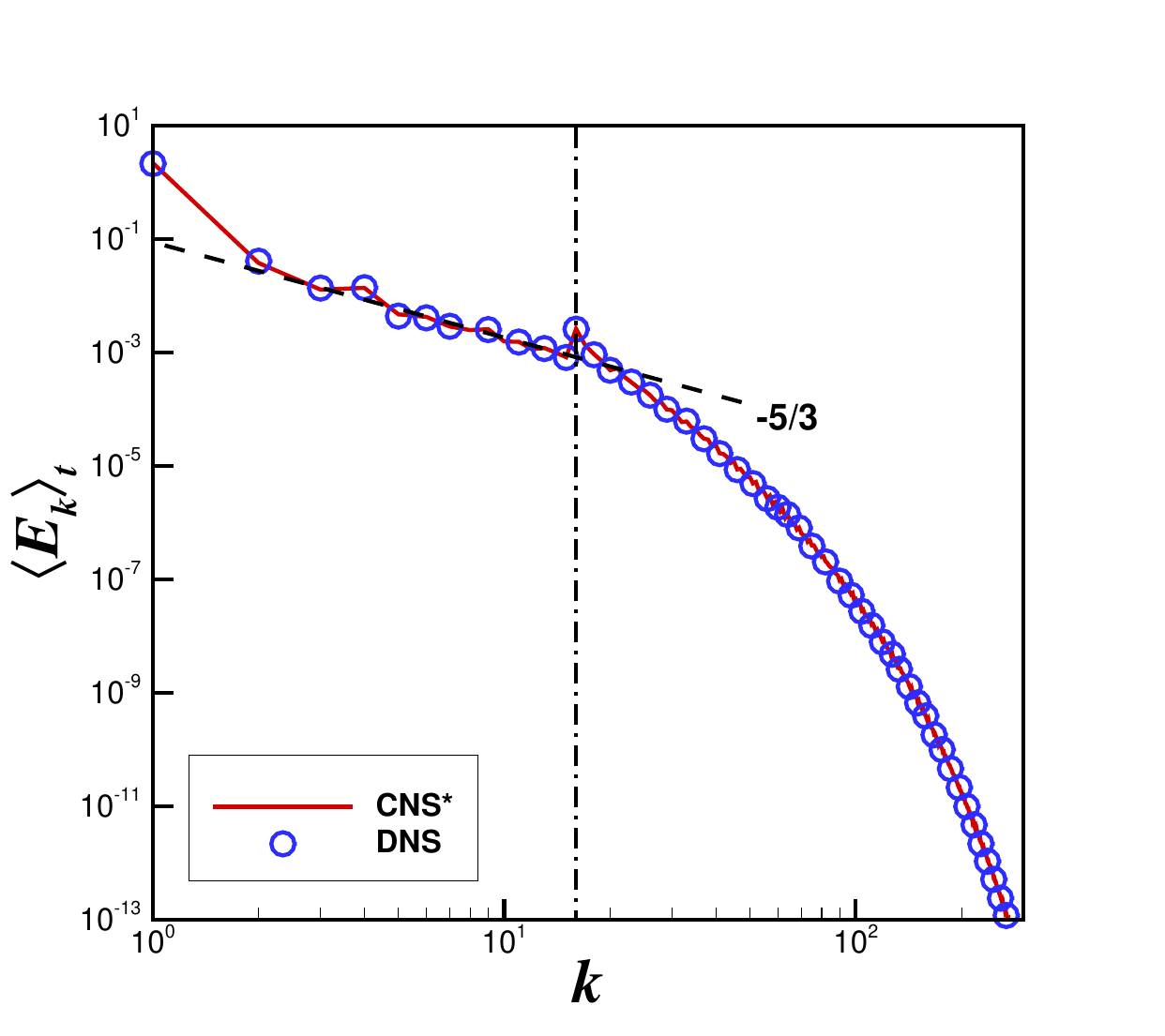}}
             \subfigure[]{\includegraphics[width=2.0in]{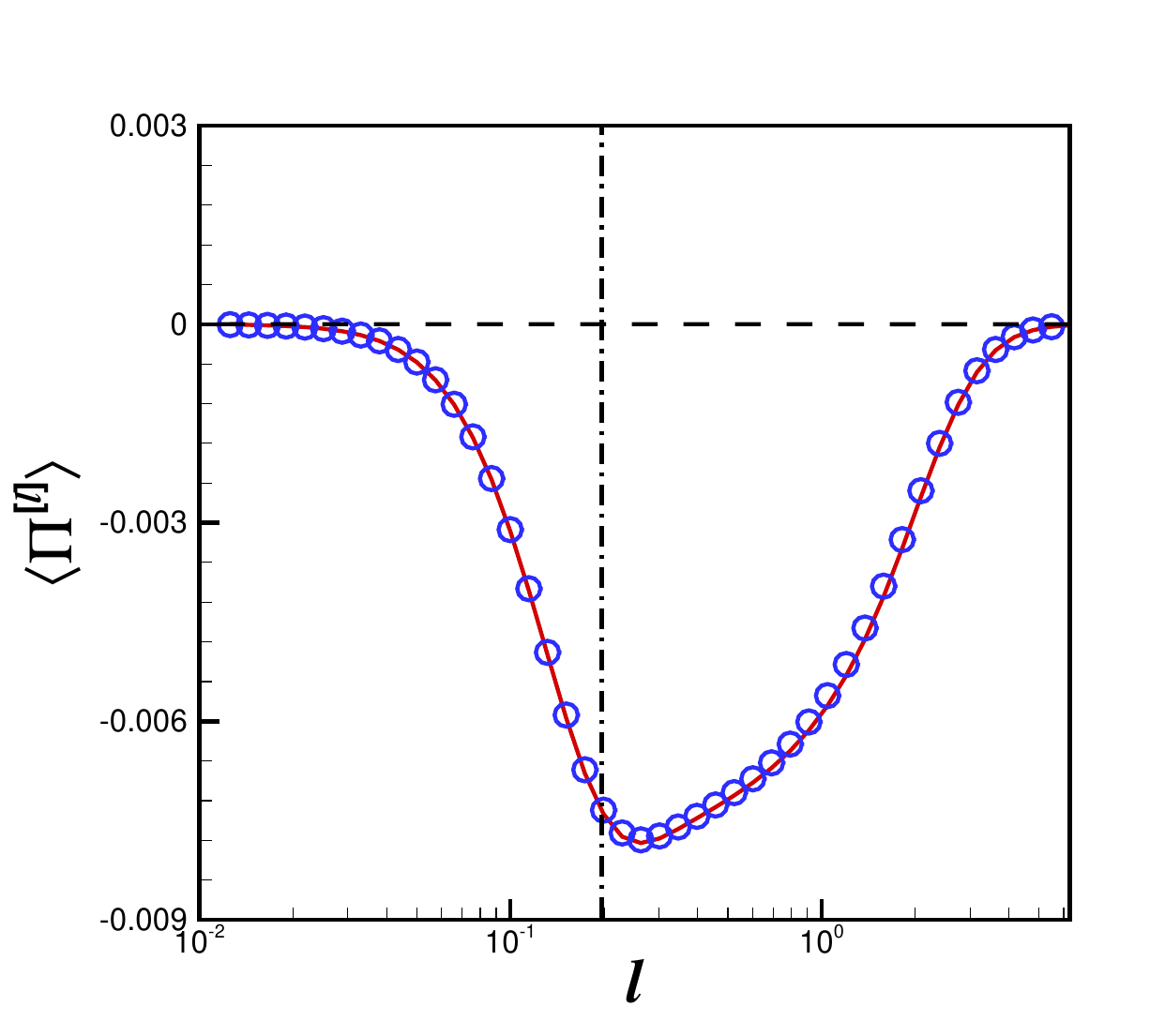}}
        \end{tabular}
    \caption{(a) Time-averaged kinetic energy spectra $\langle E_k \rangle_{t}$ of the 2D turbulent Kolmogorov flow where the black dashed line corresponds to the -5/3 power law and the black dash-dot line denotes the wave number of external force $k=n_K=16$. (b) Spatiotemporal-averaged scale-to-scale energy fluxes $\langle \Pi^{[l]} \rangle$ of the 2D turbulent Kolmogorov flow where the black dashed line denotes $\langle \Pi^{[l]} \rangle=0$ and the black dash-dot line denotes the forcing scale $l=l_f=\pi/n_K=0.196$. Red solid line is the CNS$^*$ result. Blue circles is the DNS result.}     \label{Ek_EF}
    \end{center}
\end{figure}

\begin{figure}
    \begin{center}
        \begin{tabular}{cc}
             \includegraphics[width=2in]{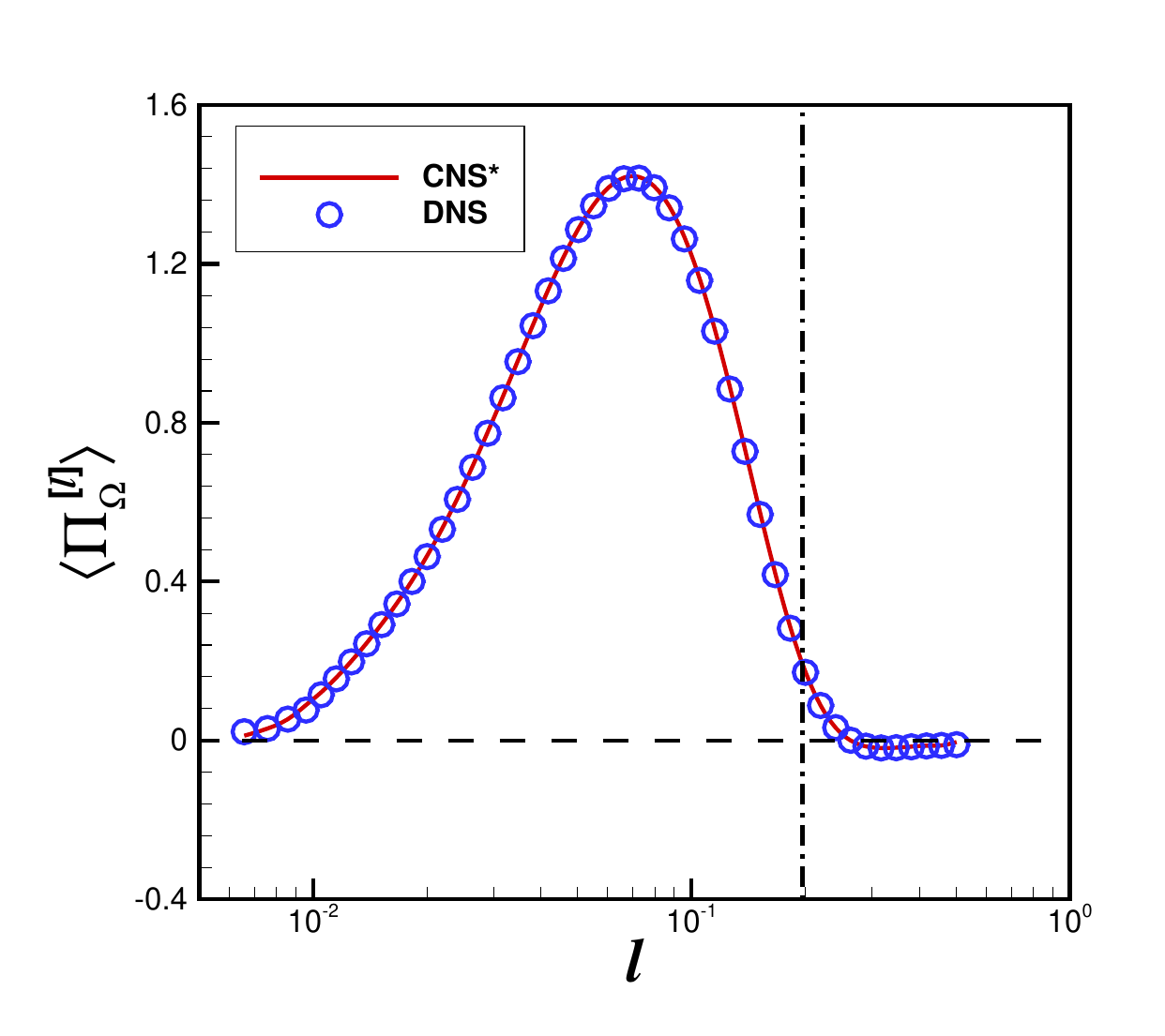}
        \end{tabular}
    \caption{Spatiotemporal-averaged scale-to-scale enstrophy fluxes $\langle \Pi_\Omega^{[l]} \rangle$ of the 2D turbulent Kolmogorov flow where the black dashed line denotes $\langle \Pi_\Omega^{[l]} \rangle=0$ \\and the black dash-dot line denotes the forcing scale $l=l_f=\pi/n_K=0.196$. \\Red solid line is the CNS$^*$ result. Blue circles display the DNS result.}     \label{EFo}
    \end{center}
\end{figure}

However,  when considering thermal fluctuation and/or stochastic environmental noise via (\ref{psi-fluctuations}) at each time step,  the time histories of the spatially averaged kinetic energy dissipation rate $\langle D\rangle_A$ as well as enstrophy dissipation rate $\langle D_{\Omega}\rangle_A$ given by CNS$^*$ and DNS are almost the {\em same}, especially for $t > 100$ which corresponds to a relatively stable state of turbulence,  as shown in Figure~\ref{D_t}.  Note that the definitions of the statistics as well as the statistic operators used in this paper are described in the appendix.  
Figure~\ref{DE-PDF} shows that the probability density functions (PDFs) of the kinetic energy dissipation rate $D(x,y,t)$ and the kinetic energy $E(x,y,t)$ given by CNS$^{*}$ and DNS agree quite well, when the integration interval of time is  $t \in [100, 300]$ corresponding to a relatively stable state of turbulence (as illustrated in Figure~\ref{D_t}). 
Similarly, the PDFs of the enstrophy dissipation rate $D_\Omega(x,y,t)$ and the enstrophy $\Omega(x,y,t)$ given by CNS$^{*}$ and DNS also agree quite well, as shown in Figure~\ref{DEo-PDF}.
Furthermore, Figure~\ref{Ek_EF} illustrates that the temporal averaged kinetic energy spectra $\langle E_k \rangle_{t}$ and the spatiotemporal-averaged scale-to-scale energy fluxes $\langle \Pi^{[l]} \rangle$ of the 2D turbulent Kolmogorov flow given by CNS$^*$ and DNS are also in accord with each other.
In addition, as shown in Figure~\ref{Ek_EF}, both the CNS$^*$ and the DNS results give the Kolmogorov's $-5/3$ power law of $\langle E_k \rangle_{t}$ when $k<n_K$, as well as the inverse energy cascade \citep{boffetta2012two, alexakis2018cascades}.
In addition, Figure~\ref{EFo} shows that the spatiotemporal-averaged scale-to-scale enstrophy fluxes $\langle \Pi_\Omega^{[l]} \rangle$ given by CNS$^*$ and DNS also agree quite well, where the direct enstrophy cascade \citep{boffetta2012two, alexakis2018cascades} exists.

\begin{figure}
    \begin{center}
        \begin{tabular}{cc}
             \subfigure[]{\includegraphics[width=2.0in]{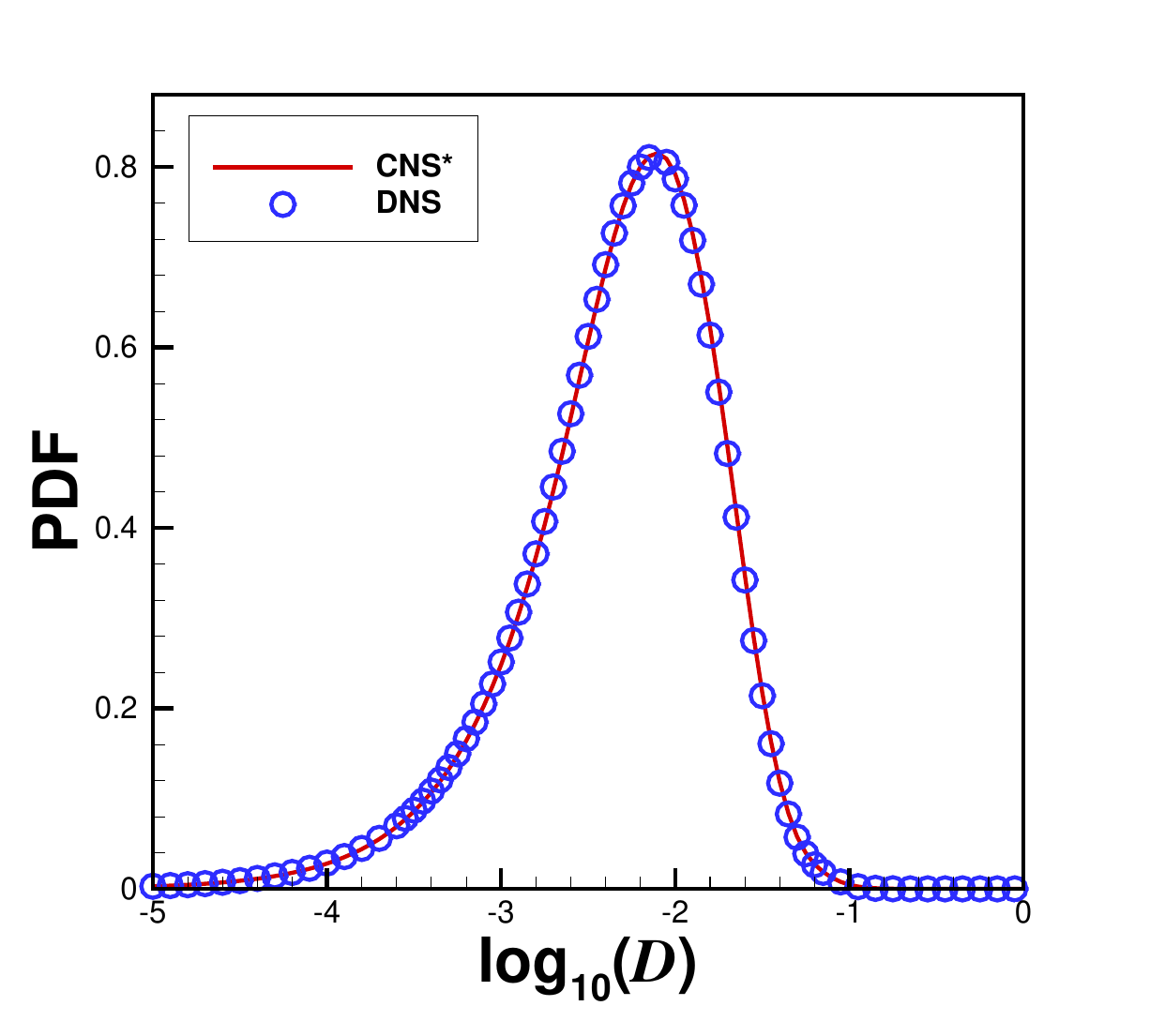}}
             \subfigure[]{\includegraphics[width=2.0in]{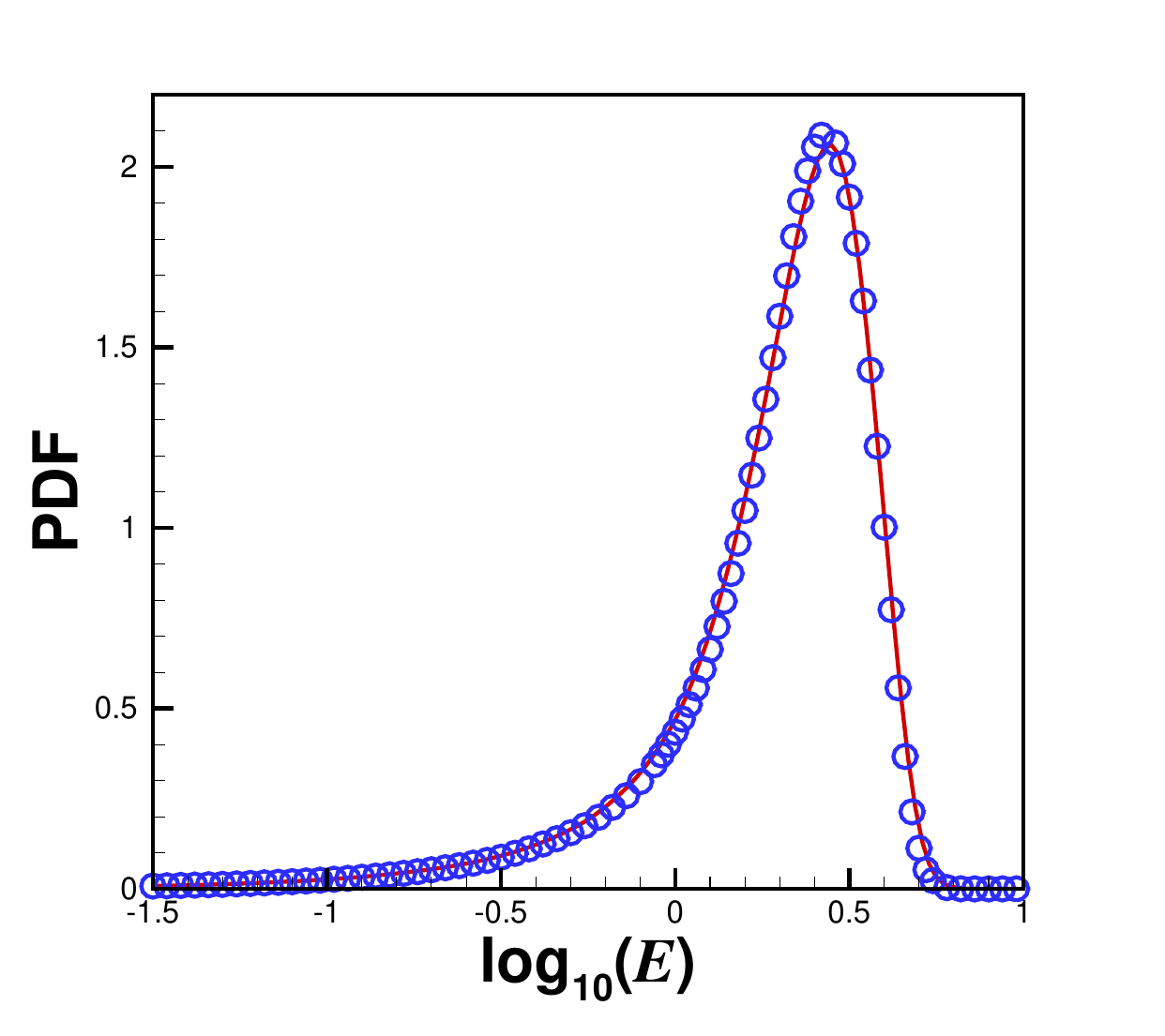}}
        \end{tabular}
    \caption{Probability density functions (PDFs) of (a) the kinetic energy dissipation rate $D(x,y,t)$ and (b) the kinetic energy $E(x,y,t)$ of the 2D turbulent Kolmogorov flow, \\where the integration is taken in $(x,y)\in[0,2\pi)^2$ and $t \in [100, 300]$: \\CNS$^*$ (red line) and DNS (blue circle).}     \label{DE-PDF}
    \end{center}
\end{figure}

\begin{figure}
    \begin{center}
        \begin{tabular}{cc}
             \subfigure[]{\includegraphics[width=2.0in]{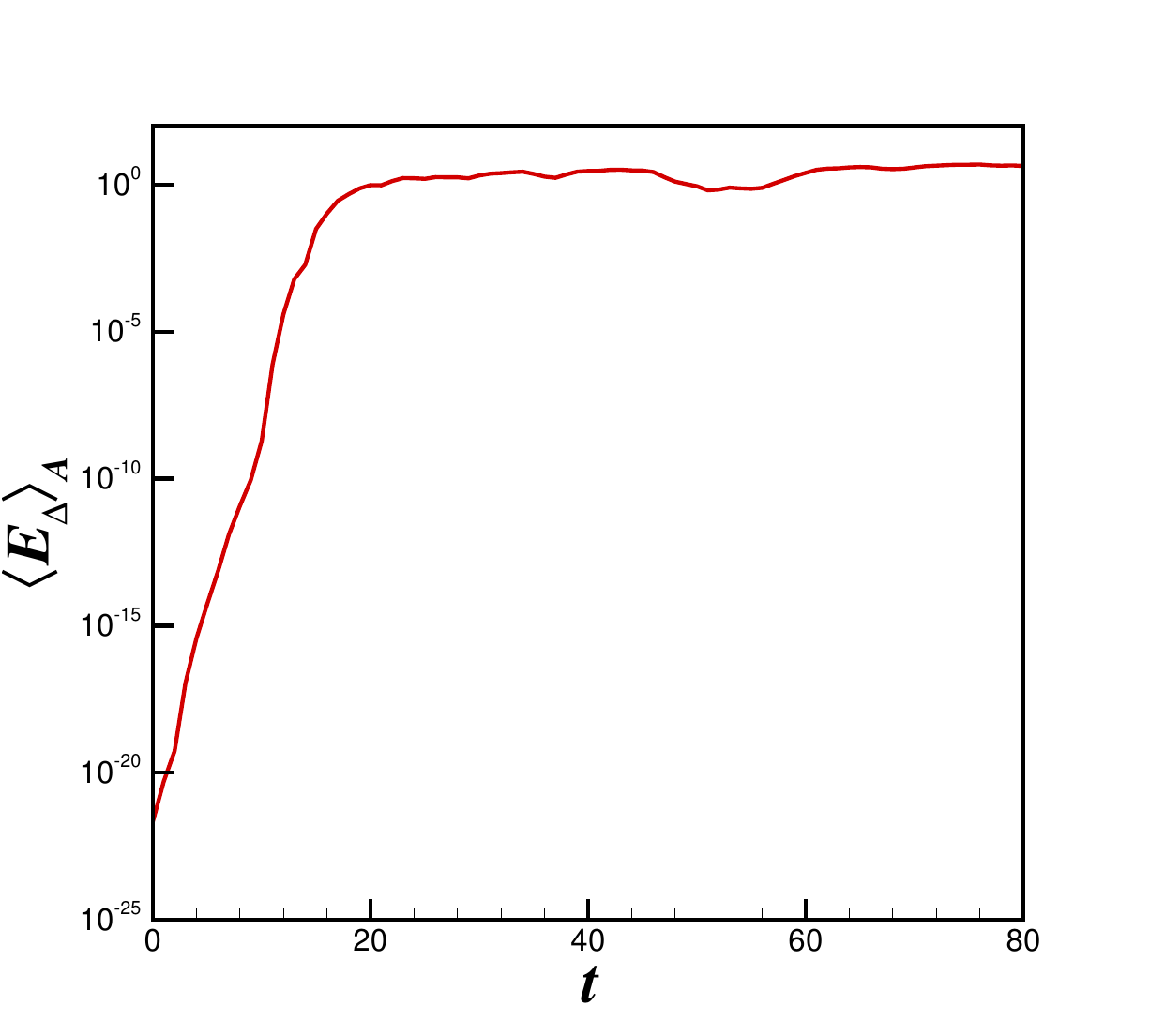}}
             \subfigure[]{\includegraphics[width=2.0in]{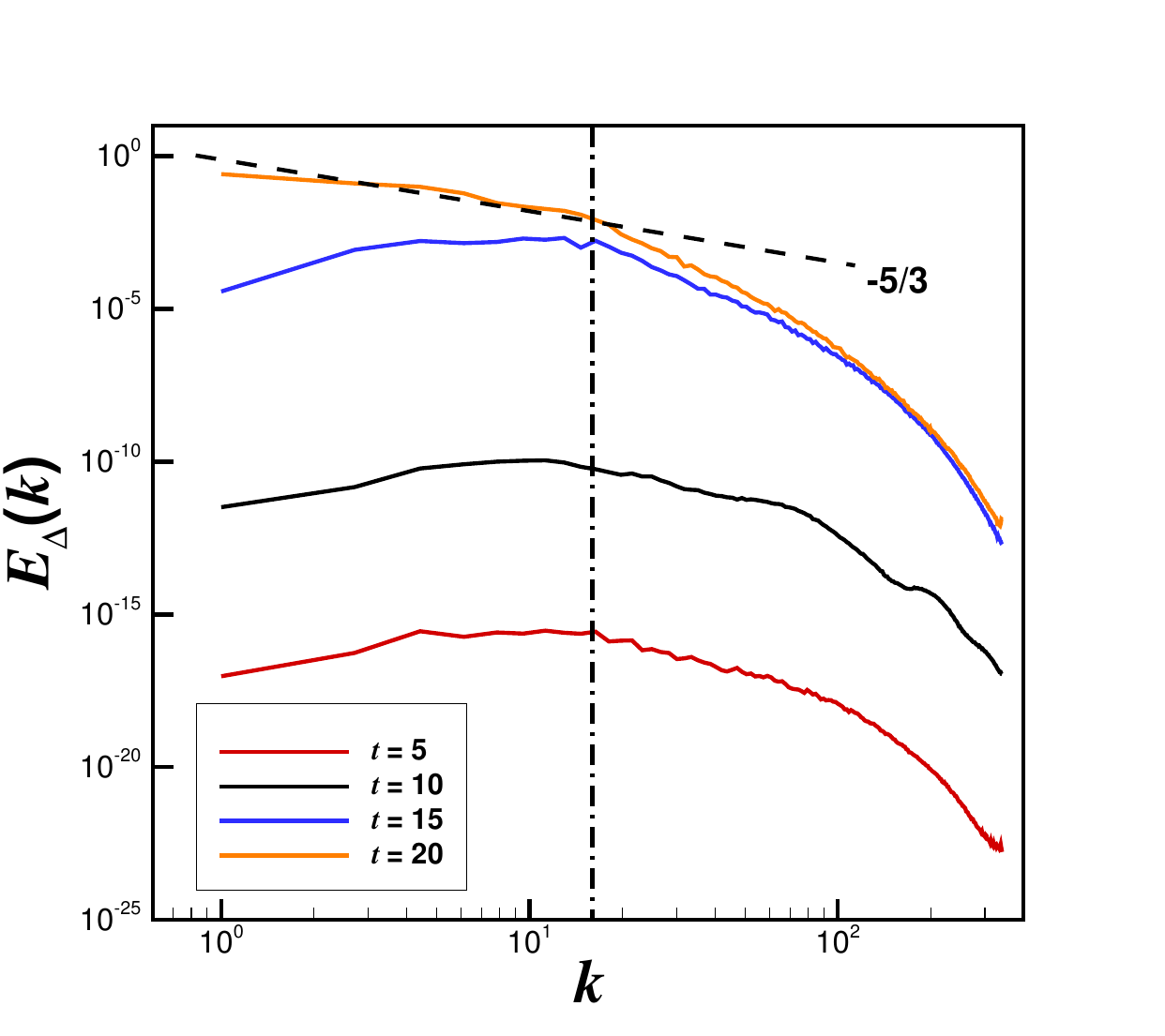}}
        \end{tabular}
    \caption{(a) Time history of the spatially averaged error/uncertainty energy $\langle E_{\Delta}\rangle_A=\langle |\Delta \mathbf{u}|^2/2 \rangle_A$. (b) Kinetic energy spectra of $\Delta \mathbf{u}$, i.e. $E_\Delta(k)$, at different times. In both (a) and (b), $\Delta \mathbf{u}=\mathbf{u}_{\mathrm{CNS}^*}-\mathbf{u}_{\mathrm{DNS}}$, where $\mathbf{u}_{\mathrm{CNS}^*}$ and $\mathbf{u}_{\mathrm{DNS}}$ correspond to the velocity fields given by CNS$^*$ and DNS, respectively.}     \label{DE}
    \end{center}
\end{figure}

For the difference between the velocity fields given by CNS$^*$ and DNS, say, $\Delta\mathbf{u}=\mathbf{u}_{\mathrm{CNS}^*}-\mathbf{u}_{\mathrm{DNS}}$, here we focus on the time evolution of the spatially averaged error/uncertainty energy $\langle E_{\Delta}\rangle_A=\langle |\Delta\mathbf{u}|^2/2 \rangle_A$ \citep{boffetta2001predictability, boffetta2017chaos, Vassilicos2023JFM}, as well as the kinetic energy spectra of $\Delta\mathbf{u}$ at different times, see Figure~\ref{DE}. 
It reveals the exponential growth of thermal fluctuation and/or stochastic environmental noise, since the thermal fluctuation and/or stochastic environmental noise added via (\ref{psi-fluctuations}) in CNS$^*$ is larger than the numerical noise in DNS.

Note that our CNS$^{*}$ result with negligible artificial numerical noise contains thermal fluctuation and/or stochastic environmental noise, but the DNS result without thermal fluctuation and/or stochastic environmental noise has rapidly become badly polluted by artificial numerical noise. 
The foregoing comparisons collectively provide evidence that artificial numerical noise in DNS is approximately  equivalent  to thermal fluctuation and/or stochastic environmental noise, at least for the 2D turbulent Kolmogorov flow considered in this paper. 
This means that the artificial numerical noise in DNS has physical significance, which provides us with a really {\em positive} perspective on artificial numerical noise in the numerical simulation of turbulence.

\subsection{Physical significance of artificial numerical noise of DNS}

\begin{figure}
    \begin{center}
        \begin{tabular}{cc}
             \subfigure[]{\includegraphics[width=2.0in]{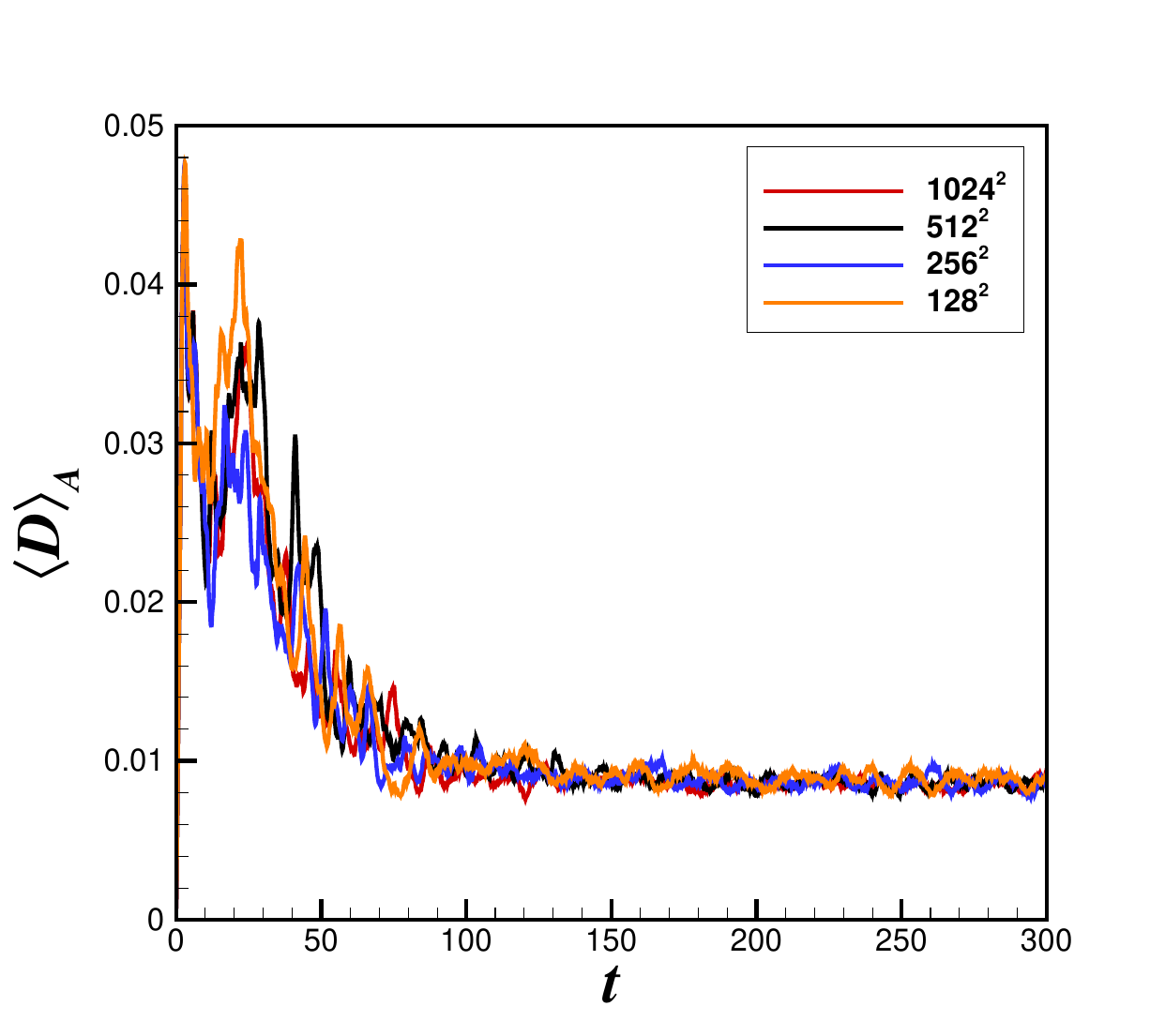}}
             \subfigure[]{\includegraphics[width=2.0in]{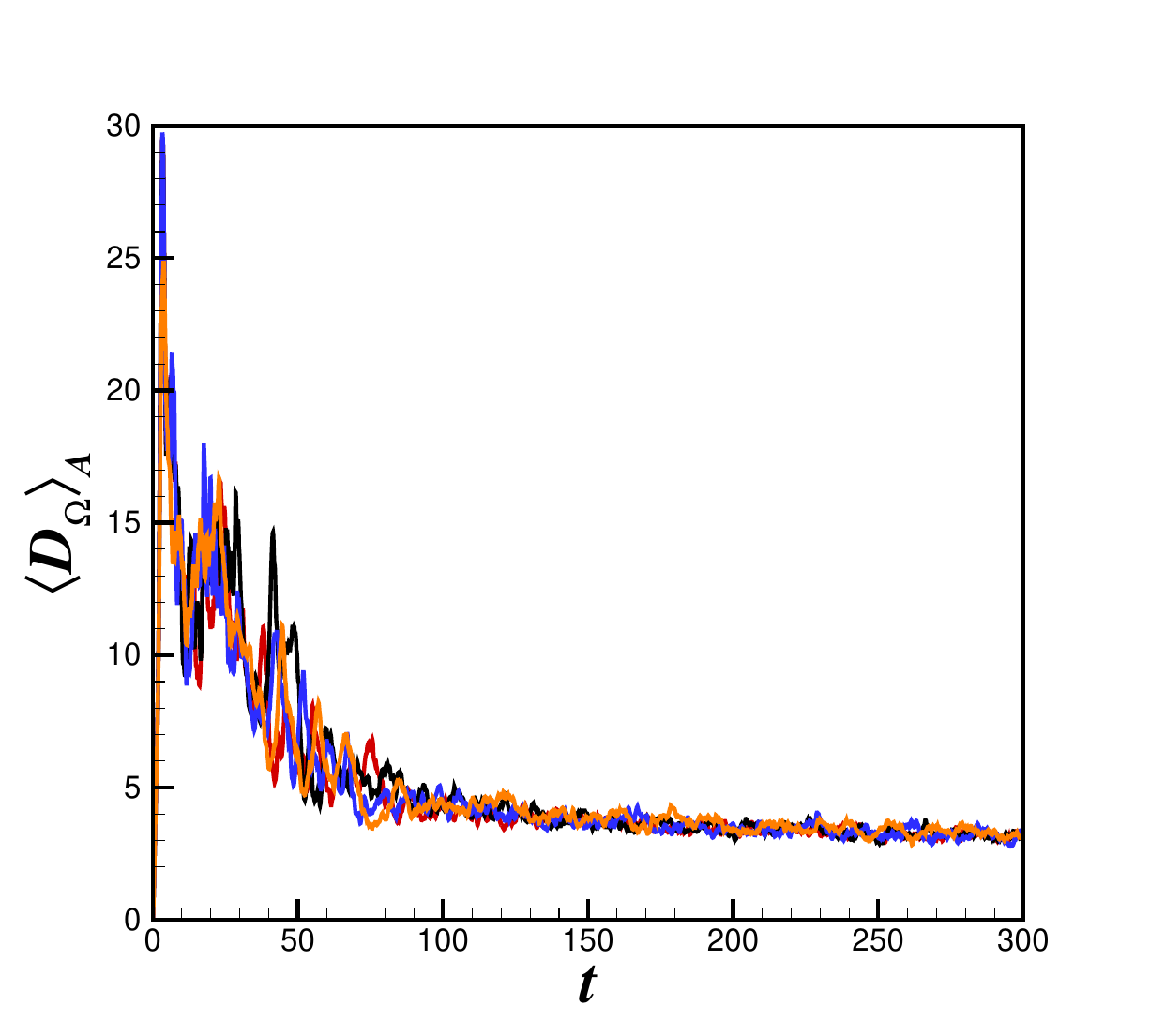}}
        \end{tabular}
    \caption{Time histories of the spatially averaged (a) kinetic energy dissipation rate $\langle D\rangle_A$ and (b) enstrophy dissipation rate $\langle D_{\Omega}\rangle_A$ of the 2D turbulent  Kolmogorov flow, given by DNS using the following four uniform meshes: $1024\times 1024$ (red line), $512\times 512$ (black line), $256\times 256$ (blue line), and $128\times 128$ (orange line).}     \label{D_t-N}
    \end{center}
\end{figure}

\begin{figure}
    \begin{center}
        \begin{tabular}{cc}
             \subfigure[]{\includegraphics[width=2.0in]{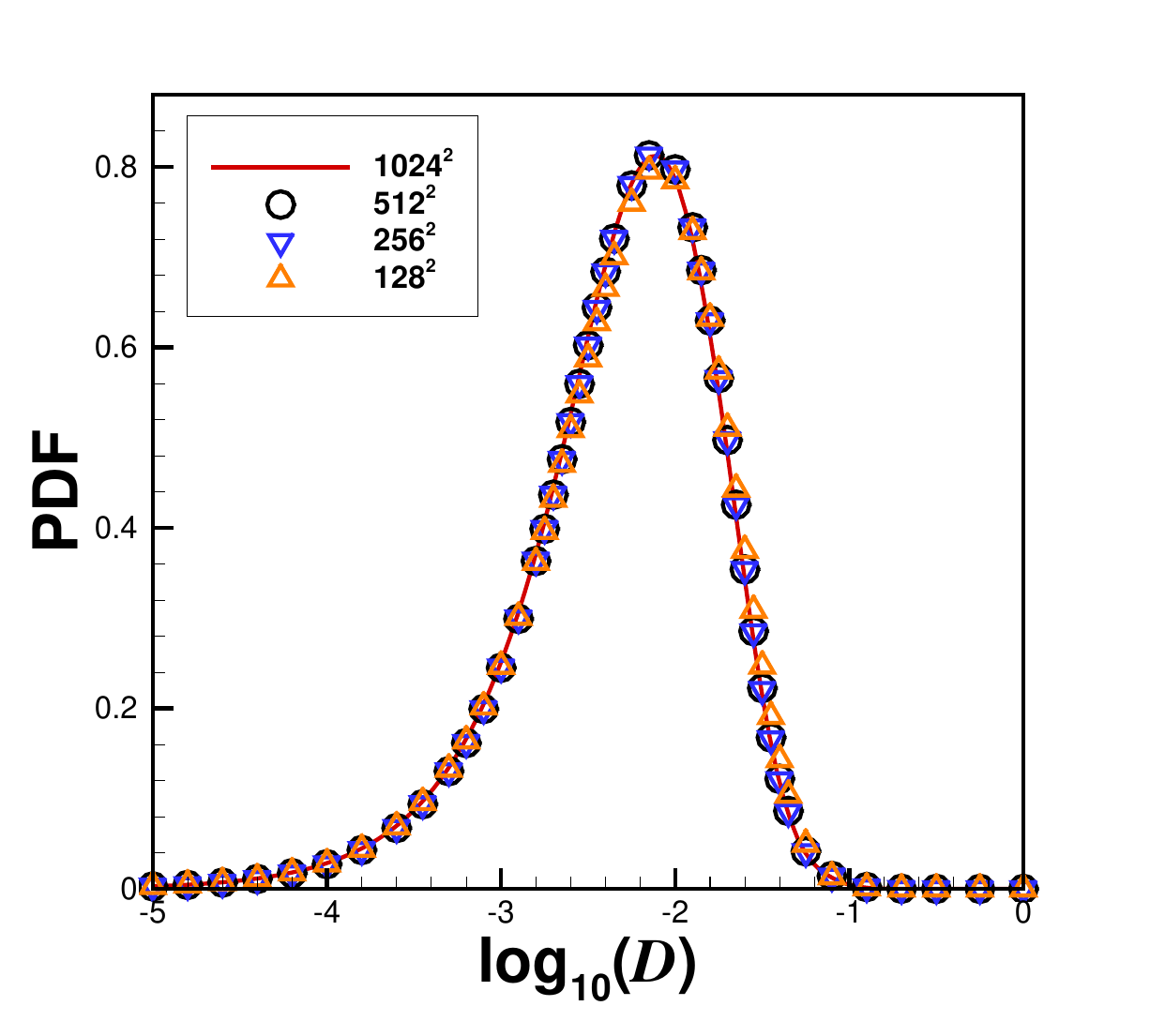}}
             \subfigure[]{\includegraphics[width=2.0in]{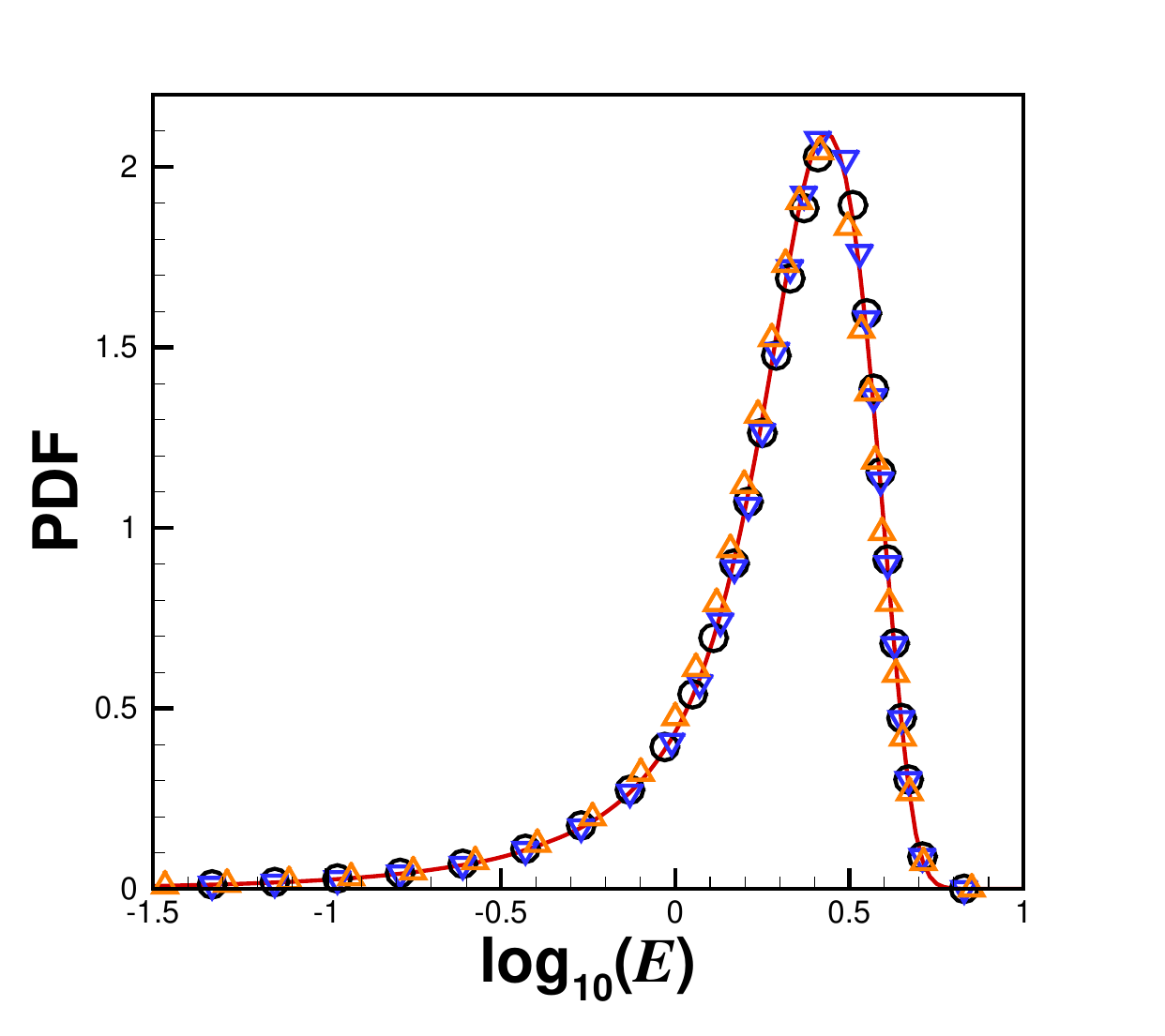}}
        \end{tabular}
    \caption{Probability density functions (PDFs) of (a) the kinetic energy dissipation rate $D(x,y,t)$ and (b) the kinetic energy $E(x,y,t)$ of the 2D turbulent Kolmogorov flow, given by DNS using the following four uniform meshes: $1024\times 1024$ (red line), $512\times 512$ (black circle), $256\times 256$ (blue inverted triangle), and $128\times 128$ (orange triangle).}     \label{DE-PDF-N}
    \end{center}
\end{figure}

\begin{figure}
    \begin{center}
        \begin{tabular}{cc}
             \subfigure[]{\includegraphics[width=2.0in]{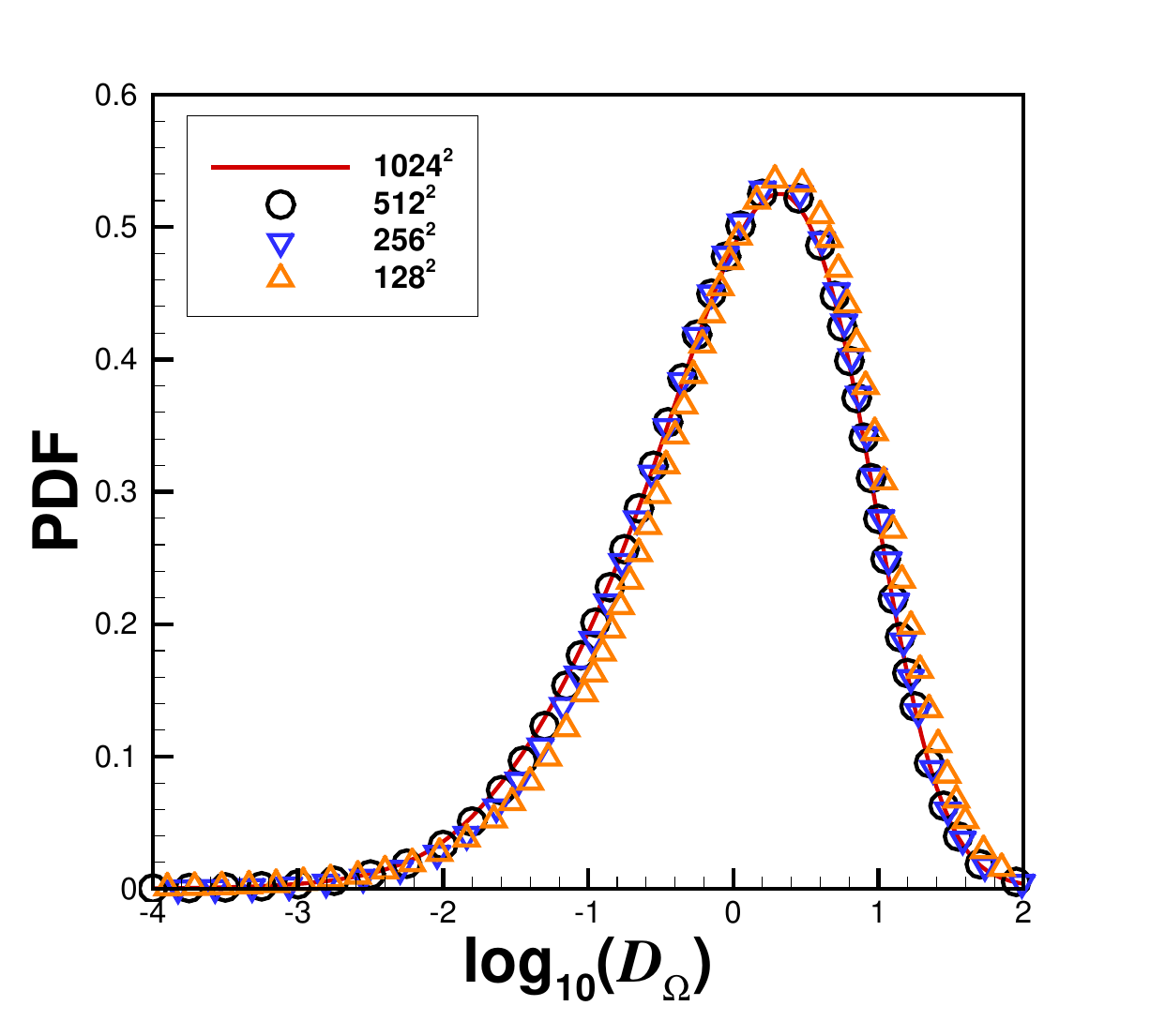}}
             \subfigure[]{\includegraphics[width=2.0in]{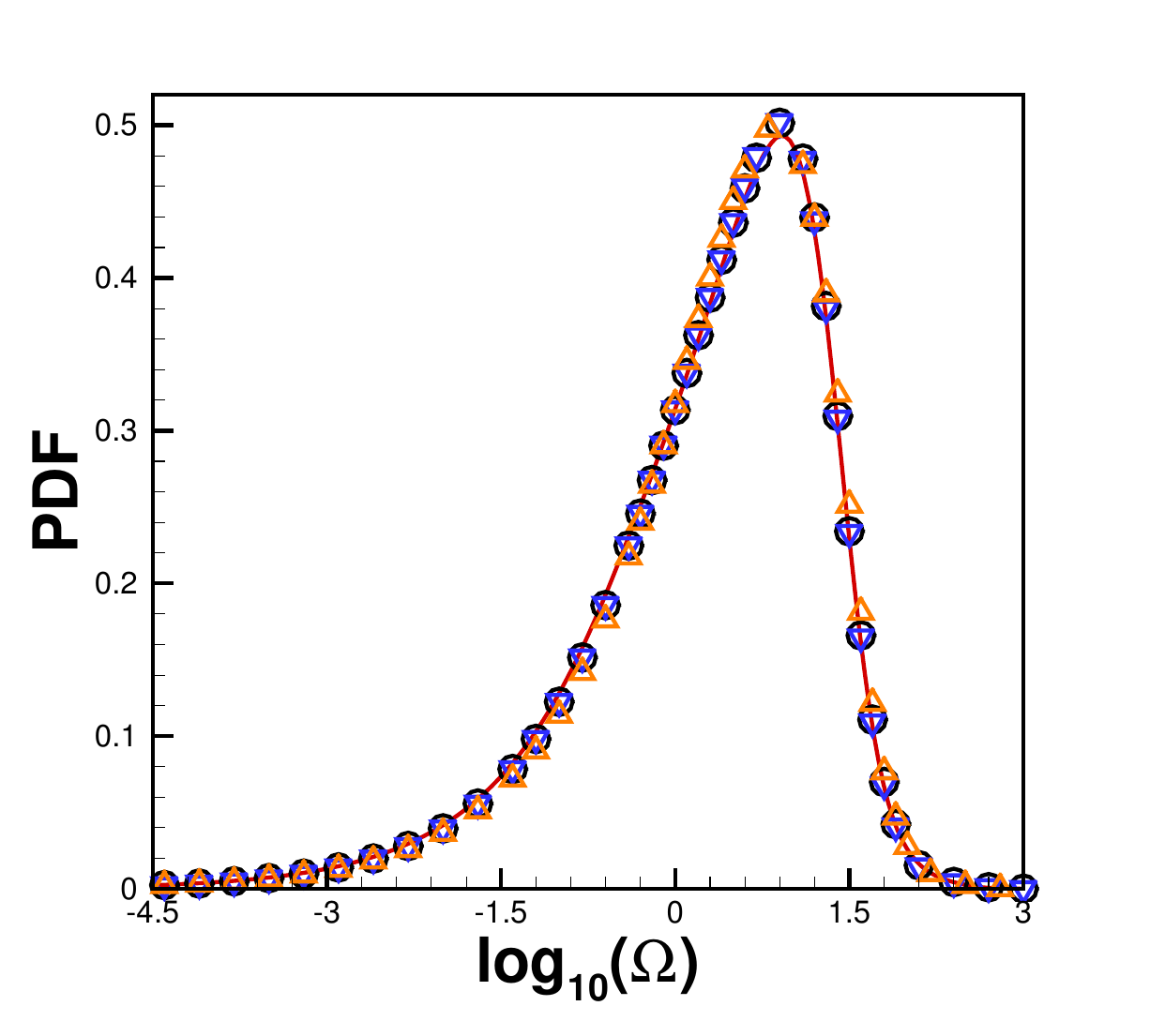}}
        \end{tabular}
    \caption{Probability density functions (PDFs) of (a) the enstrophy dissipation rate $D_\Omega(x,y,t)$ and (b) the enstrophy $\Omega(x,y,t)$ of the 2D turbulent Kolmogorov flow, given by DNS using the following four uniform meshes: $1024\times 1024$ (red line), $512\times 512$ (black circle), $256\times 256$ (blue inverted triangle), and $128\times 128$ (orange triangle).}     \label{DEo-PDF-N}
    \end{center}
\end{figure}

Given that the artificial numerical noise in DNS is approximately equivalent to thermal fluctuation and/or stochastic environmental noise, artificial numerical noise can thus be regarded from a totally different physical perspective: different sources of artificial numerical noise in DNS, arising from different algorithms, different spatial meshes, different time steps, etc., correspond to different thermal fluctuation and/or stochastic environmental noise. 
From this physical viewpoint of artificial numerical noise, DNS results given by various numerical algorithms with different levels of {\em artificial} numerical noise correspond to different turbulent flows under different levels of {\em physical} disturbances such as thermal fluctuation and/or  stochastic environmental noise: all of them could be correct and have physical meaning.   

\begin{figure}
    \begin{center}
        \begin{tabular}{cc}
             \subfigure[]{\includegraphics[width=2.0in]{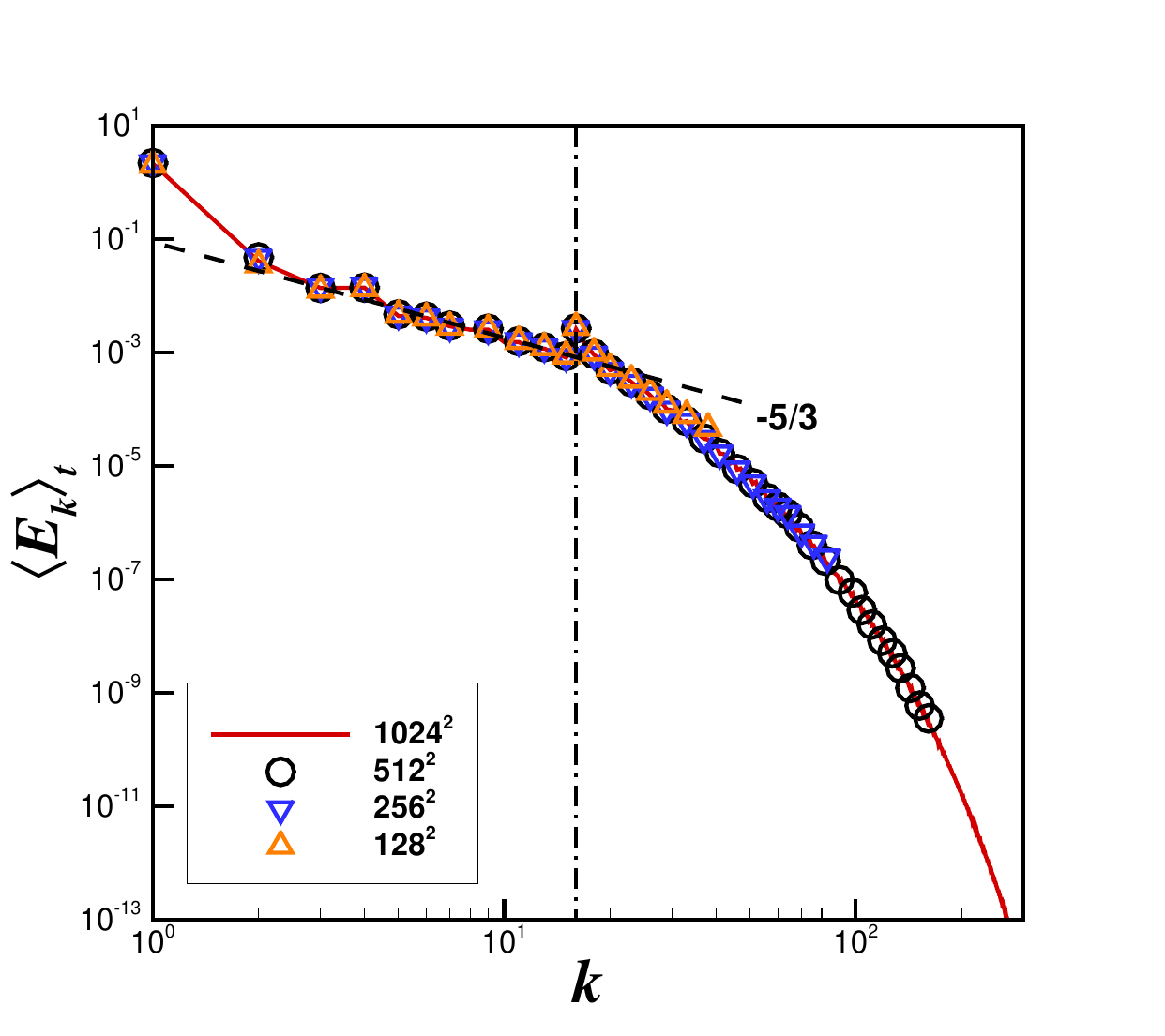}}
             \subfigure[]{\includegraphics[width=2.0in]{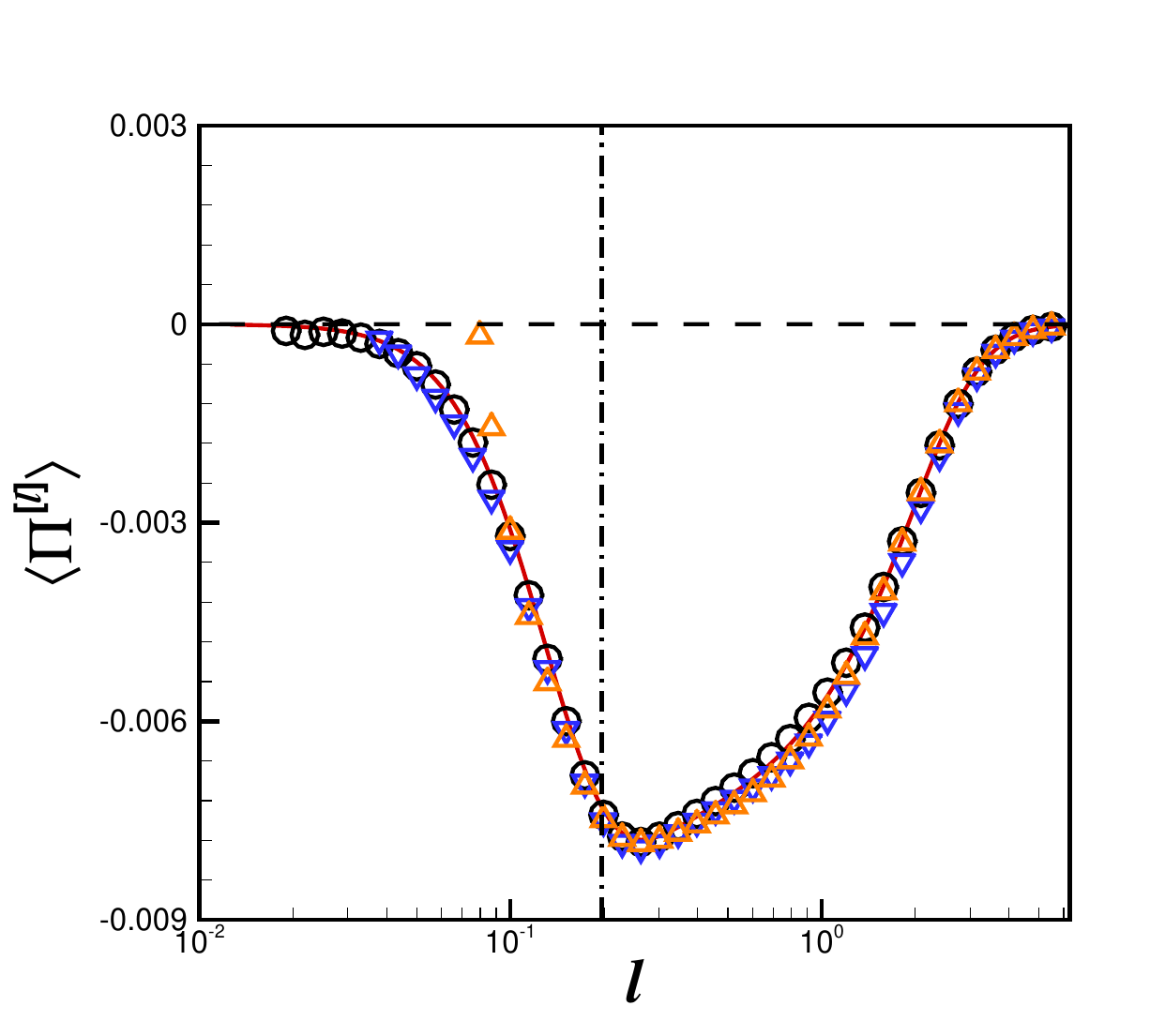}}
        \end{tabular}
    \caption{(a) Time-averaged kinetic energy spectra $\langle E_k \rangle_{t}$ of the 2D turbulent Kolmogorov flow. The black dashed line corresponds to the -5/3 power law and the black dash-dot line denotes the wave number of external force $k=n_K=16$. (b) Spatiotemporal-averaged scale-to-scale energy fluxes $\langle \Pi^{[l]} \rangle$ of the 2D turbulent Kolmogorov flow where the black dashed line denotes $\langle \Pi^{[l]} \rangle=0$ and the black dash-dot line denotes the forcing scale $l=l_f=\pi/n_K=0.196$. In both (a) and (b), the results were obtained using DNS on $1024\times 1024$ (red solid line), $512\times 512$ (black circle), $256\times 256$ (blue inverted triangle), and $128\times 128$ (orange triangle) uniform meshes.}     \label{Ek_EF-N}
    \end{center}
\end{figure}

\begin{figure}
    \begin{center}
        \begin{tabular}{cc}
             \includegraphics[width=2.3in]{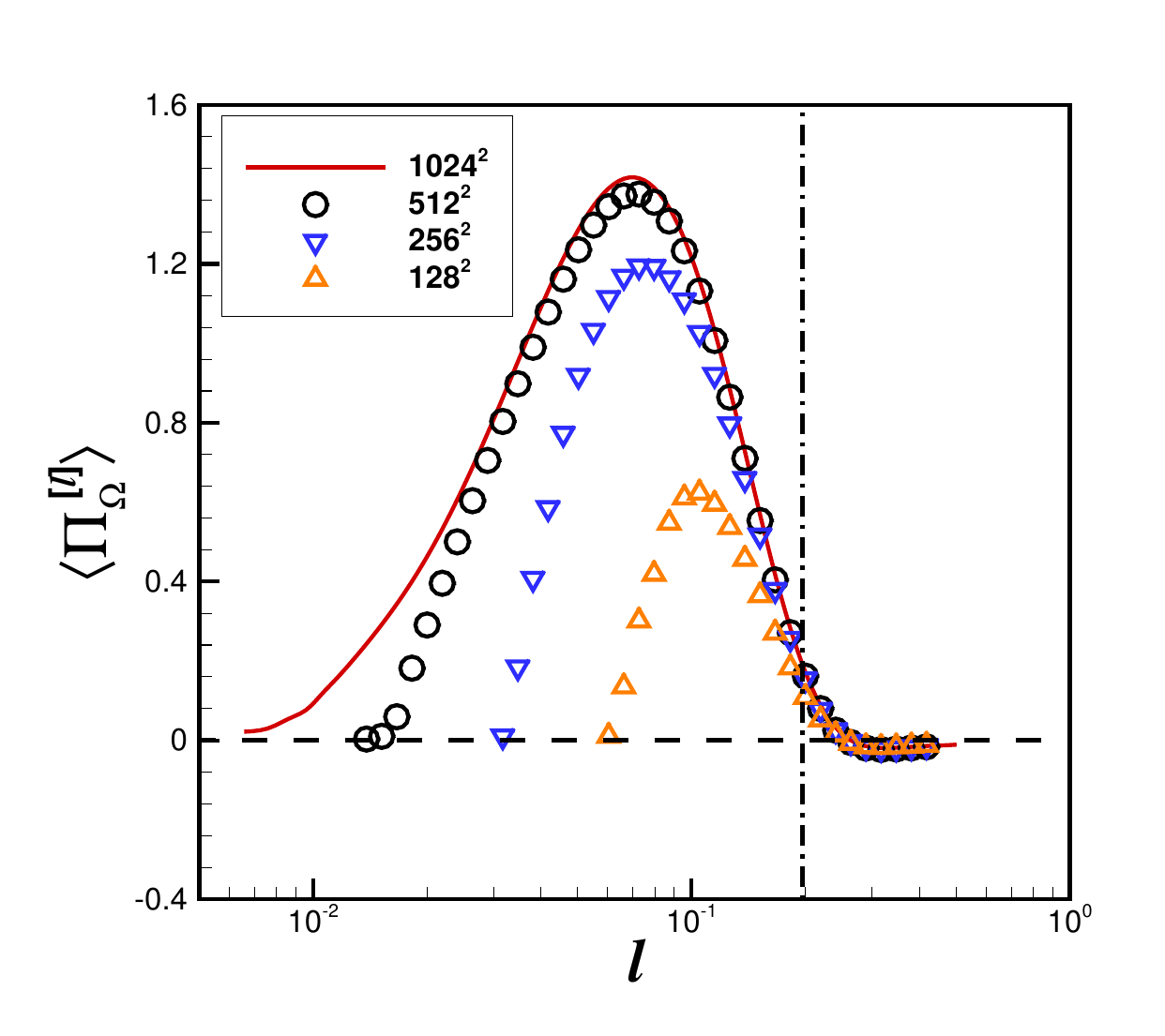}
        \end{tabular}
    \caption{Spatiotemporal-averaged scale-to-scale enstrophy fluxes $\langle \Pi_\Omega^{[l]} \rangle$ of the 2D turbulent Kolmogorov flow, given by DNS using the following four uniform meshes: $1024\times 1024$ (red line), $512\times 512$ (black circle), $256\times 256$ (blue inverted triangle), and $128\times 128$ (orange triangle), where the black dashed line denotes $\langle \Pi_\Omega^{[l]} \rangle=0$ and the black dash-dot line denotes the forcing scale $l=l_f=\pi/n_K=0.196$.}     \label{EFo-N}
    \end{center}
\end{figure}

For example, we can similarly obtain {\em different} DNS results by means of the {\em same} strategy as mentioned above, i.e. 4th-order Runge-Kutta method with time-step $\Delta t=10^{-4}$ in double precision, as well as the {\em same} pseudo-spectral method in space but using the three {\em different} uniform meshes, i.e. $512\times 512$, $256\times 256$, and $128\times 128$, corresponding to {\em different} levels of spatial truncation error.  
As shown in Figure~\ref{D_t-N}, the time histories of the spatially averaged kinetic energy dissipation rate $\langle D\rangle_A$ and enstrophy dissipation rate $\langle D_{\Omega}\rangle_A$ given by DNS using the above-mentioned three {\em different} uniform meshes are almost the {\em same} as those given by DNS using the finest uniform mesh $1024\times 1024$, especially when $t>100$ corresponding to a relatively stable state of turbulence.
Besides, the PDFs of kinetic energy dissipation rate $D(x,y,t)$ and kinetic energy $E(x,y,t)$ are also almost the same, as shown in Figure~\ref{DE-PDF-N} (a) and (b), respectively.
Similarly, the PDFs of the enstrophy dissipation rate $D_\Omega(x,y,t)$ and the enstrophy $\Omega(x,y,t)$ given by different uniform meshes also agree quite well, as shown in Figure~\ref{DEo-PDF-N}.
As shown in Figure~\ref{Ek_EF-N} (a), there exists no obvious difference between the temporal averaged kinetic energy spectra $\langle E_k \rangle_{t}$ obtained via the four meshes: {\em all} satisfy the Kolmogorov $-5/3$ power law. Figure~\ref{Ek_EF-N} (b) shows that the spatiotemporal-averaged scale-to-scale energy fluxes $\langle \Pi^{[l]} \rangle$ obtained via the different meshes agree well with each other mostly, except at $l\approx10^{-1}$ for the uniform $128\times 128$ mesh that is too sparse to describe accurately the small-scale turbulent flow in detail. 
However, even so, all of them correctly lead to the physical conclusion that the energy cascade is inverse, i.e. directed from small-scale to large-scale.
In addition, all the spatiotemporal-averaged scale-to-scale enstrophy fluxes $\langle \Pi_\Omega^{[l]} \rangle$ given by these different uniform meshes display the direct enstrophy cascade, as shown in  Figure~\ref{EFo-N}.
The foregoing indicate that the statistical results given by DNS using the four uniform meshes of different resolution agree quite well for the 2D turbulent Kolmogorov flow under consideration. This is indeed a very surprising result.  

Traditionally, DNS has been widely regarded as providing `reliable' benchmark solutions of turbulence as long as the grid spacing is fine enough, say, less than the enstrophy dissipative scale for 2D turbulence \citep{boffetta2012two}, and the time-step is sufficiently small, say, satisfying the Courant-Friedrichs-Lewy condition, i.e. Courant number $<1$  \citep{Courant1928}. In principle these two conditions {\em must} be satisfied for DNS of turbulence.        
As shown in Figure~\ref{D_t-N}(b), all spatially averaged enstrophy dissipation rates $\langle D_{\Omega}\rangle_A(t)$ given by DNS using the four different meshes are almost the same when $t>100$. Thus, integrated over $(x,y)\in[0,2\pi)^2$ and $t \in [100, 300]$, we obtain almost the same spatiotemporal-averaged enstrophy dissipation rates $\langle D_{\Omega}\rangle=3.5$ for all DNS results using the four different uniform meshes, and the corresponding enstrophy dissipative scale \citep{boffetta2010evidence} is  
\begin{equation}
\langle \eta_{\Omega} \rangle \approx Re^{-1/2}\langle D_{\Omega} \rangle^{-1/6}=0.018.    \label{scale}
\end{equation}
Thus, we have the corresponding grid spacing $\Delta_{1024}=2\pi/1024\approx0.34\langle \eta_{\Omega} \rangle$ for $1024\times 1024$ mesh, $\Delta_{512}\approx0.68\langle \eta_{\Omega} \rangle$ for $512\times 512$ mesh, $\Delta_{256}\approx1.36\langle \eta_{\Omega} \rangle$ for $256 \times 256$ mesh, and $\Delta_{128}\approx2.73\langle \eta_{\Omega} \rangle$ for $128\times 128$ mesh, respectively. It should be emphasized that, although the grid spacing is fine enough only for the two uniform meshes $1024\times 1024$ and $512\times 512$, {\em all} the statistical results obtained using DNS on the {\em different} meshes agree quite well with each other, even if the grid spacing $\Delta_{128}$ is even 2.73 times larger than the enstrophy dissipative scale $\langle \eta_{\Omega} \rangle$.  It is hard to explain this kind of agreement in the traditional frame of the DNS.

However, the above-mentioned phenomena can be fully explained by considering the physical meaning of artificial numerical noise of DNS, revealed in \S~3.1. Note that the DNS algorithms using the four different uniform meshes have different levels of artificial numerical noise, which are approximately equivalent to different levels of thermal fluctuation and/or stochastic environmental noise. Therefore, each DNS result corresponds to a 2D turbulent Kolmogorov flow under a particular level of thermal fluctuation and/or stochastic environmental noise: their spatiotemporal trajectories are certainly different, but all of them  have physical meaning. In other words, all are physically correct! 
So, it is {\em meaningless} to try and say which one among them is better given that {\em all} of them are correct in physics, corresponding to a turbulent flow under a kind of thermal fluctuation and/or stochastic environmental noise.      

Note that, for the 2D turbulent Kolmogorov flow under consideration, all the statistical results given by DNS using the four different uniform meshes (and even  different time-steps) agree well, indicating that {\em statistical stability} has been achieved and the simulations are insensitive to different levels of disturbance.
It should be emphasized that, for turbulent flow that is statistically stabile, one can use even a sparse $128 \times 128$ mesh (i.e. requiring much {\em less} CPU time) to gain almost the {\em same} statistical results as those on the finest $1024\times 1024$ mesh. Thus, statistical stability is very important for numerical simulation of turbulence in practice.  
This is exactly the reason why \citet{Liao2023book} proposed the so-called  `{\em modified fourth Clay millennium problem}':
\begin{itemize}
\item[]  ``The existence, smoothness and {\em statistic stability} of the Navier-Stokes equation: Can we prove the existence, smoothness and {\em statistic  stability} (or {\em instability}) of the solution of the Navier-Stokes equation with physically proper boundary and initial conditions?''
\end{itemize}
Unfortunately, such kind of statistic stability does {\em not} always exist for all turbulent flows, as illustrated below.       

\subsection{An example of turbulence with statistic instability}

\begin{figure}
    \begin{center}
        \begin{tabular}{cc}
             \includegraphics[width=3.4in]{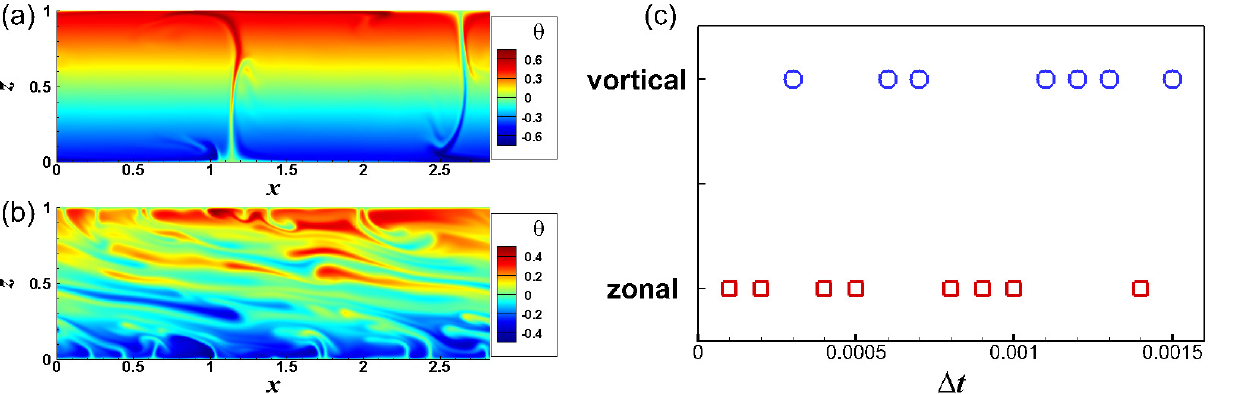}
        \end{tabular}
    \caption{(a)-(b) $\theta$ (temperature departure from a linear variation background) at time $t=250$ of 2D turbulent Rayleigh-B\'{e}nard convection (RBC) for Rayleigh number $Ra=5\times10^7$, Prandtl number $Pr=6.8$ and aspect ratio $\Gamma=2\sqrt{2}$, given by DNS with different time steps: (a) non-shearing vortical/roll-like flow given by $\Delta t=1.1\times10^{-3}$ and (b)  zonal flow given by $\Delta t=10^{-3}$. (c) Final flow type of the turbulent RBC versus time-step $\Delta t$ of DNS for the same Rayleigh-B\'{e}nard convection: either non-shearing vortical/roll-like flow (blue circle) or zonal flow (red square).}     \label{RB}
    \end{center}
\end{figure}

Let us recall that, by means of CNS, \cite{Qin2022JFM} investigated the large-scale influence of numerical noise as tiny artificial stochastic disturbances on sustained turbulence, i.e. 2D turbulent Rayleigh-B\'{e}nard convection (RBC) for aspect ratio $\Gamma = 2\sqrt{2}$, Prandtl number $Pr$ = 6.8 (corresponding to water at room temperature, 20\,$^{\circ}$C), and Rayleigh number $Ra = 6.8 \times 10^{8}$ (corresponding to a turbulent state). It was found \citep{Qin2022JFM}  that the CNS benchmark solution always sustains a non-shearing vortical/roll-like convection throughout the whole simulation, however the DNS result is a kind of vortical/roll-like convection at the beginning but finally turns into a kind of zonal flow.  The two distinct types of turbulent convection are also confirmed  by \cite{wang2023lifetimes}. 
This illustrated that numerical noise as tiny artificial stochastic disturbance could lead to the simulated turbulence experiencing large-scale deviations not only in spatiotemporal trajectories but also even in statistics and type of flow. This is a good example of turbulence having statistic instability.  

To show such kind of statistic instability in more detail, let us further consider here the same 2D turbulent Rayleigh-B\'{e}nard convection (RBC), governed by the same mathematical equations subject to the same initial/boundary conditions with the same physical parameters as those used by \cite{Qin2022JFM}, except for a smaller Rayleigh number $Ra=5\times10^7$.   The same DNS algorithm using the same uniform mesh $1024\times 1024$ as that by \cite{Qin2022JFM} is adapted but with {\em various} time-steps, which correspond to different levels of artificial numerical noise that are approximately equivalent to different levels of thermal fluctuation and/or stochastic environmental noise, as verified in \S~3.1 of the present paper.             

It is found that the flow type and the corresponding statistical results given by DNS  are rather sensitive to the value of time-step $\Delta t$. For example,  
the time-step $\Delta t=1.1\times10^{-3}$ gives a non-shearing vortical/roll-like flow, but the time-step $\Delta t=10^{-3}$ corresponds to a zonal flow, as shown in Figure~\ref{RB} (a) and (b). It should be emphasized that the difference between the two time-steps is merely $10^{-4}$. Note that the final flow type of the 2D  turbulent Rayleigh-B\'{e}nard convection is rather sensitive to the time-step $\Delta t$ of DNS, as shown in Figure~\ref{RB}(c): as  $\Delta t$ varies from $10^{-4}$ to $1.5\times10^{-3}$ with uniform interval $10^{-4}$, where the final flow type consistently fluctuates between non-shearing vortical/roll-like flow and zonal flow, which of course causes the statistics of the corresponding turbulent flow to fluctuate, i.e. promoting statistical instability.

Note that artificial numerical noise of DNS is approximately equivalent to thermal fluctuation and/or stochastic environmental noise, as verified in \S~3.1. 
Therefore, each of our DNS results given by different time-step corresponds to a kind of turbulent flow under a different level of thermal fluctuation and/or stochastic environmental noise (regardless of its different statistics), and thus has proper physical meaning, no matter whether the flow type is non-shearing vortical/roll-like flow or zonal flow.  

\section{Concluding remarks and discussions}

Using clean numerical simulation (CNS), we provide rigorous evidence that, for DNS of turbulence, artificial numerical noise is approximately equivalent to thermal fluctuation and/or stochastic environmental noise. 
This reveals the physical significance of artificial numerical noise in DNS of  turbulence governed by the Navier-Stokes equations.
In other words, the results produced by DNS on different numerical meshes should correspond to turbulent flows under different levels of internal/external physical disturbances.  
More importantly,  all could have physical meaning even if their statistics are quite different, so long as these equivalent disturbances are reasonable and practical  in physics.  This provides a positive perspective on artificial numerical noise in DNS of turbulence.         

Note that, by means of DNS itself, it is impossible to verify that artificial numerical noise in DNS is approximately equivalent to thermal fluctuation and/or stochastic environmental noise, because the DNS result rapidly becomes badly polluted by its inherent numerical noise.
This illustrates that CNS, whose artificial numerical noise is negligible over a finite, sufficiently long time interval, can indeed provide us with a useful tool by which to investigate the propagation and evolution of artificial/physical micro-level disturbances and their large-scale influence on turbulence.        

Similarly, due to the butterfly-effect of chaos, artificial numerical noise will enlarge quickly to macro-level of a chaotic system. Thus, the foregoing conclusions should also hold in general for a chaotic system; in other words, the artificial numerical noise of every chaos should be equivalent to its physical and/or stochastic environmental noise.

Data-driven artificial intelligence (AI) and machine learning (ML) have been widely used in fluid mechanics.  Note that  all data contain noise and all algorithms in ML and/or AI are likely  to introduce some artificial noises.  Obviously, the physical significance of artificial numerical noise in DNS could provide a new viewpoint for data-driven AI and ML in fluid mechanics.     

It is a well-known phenomenon in computational fluid dynamics (CFD) that numerical simulations of a turbulent flow given by various algorithms are quite different from each other and from the corresponding physical experiment when its experimental result has {\em not} been announced, but generally {\em agree well} with the experimental result as soon as it is announced. The conventional explanation for this `famous' phenomenon is that those simulations that exhibit obvious deviations from the physical experimental results must be wrong, primarily because the numerical algorithm and/or spatial mesh are simply not good enough, leading to too high a level of artificial numerical noise to obtain the `correct' numerical results.
However, according to the new viewpoint about artificial numerical noise revealed in this paper, many (or even all) of these numerical results might be correct in physics, even if there exist huge deviations between them (as illustrated in \S~3.3, see Figure~\ref{RB}), because their internal/external physical disturbances might be quite {\em different} but the corresponding physical experiment simply corresponds to {\em one} special case of physical disturbance. Hopefully, the physical significance of artificial numerical noise as a really positive viewpoint revealed in this paper could be of benefit to greatly deepen our understanding about the so-called `crisis of reproducibility'  for CFD \citep{Baker2016Nature}.

Note that, if thermal fluctuation and/or stochastic environmental noise is {\em not} considered, the CNS result retains the same spatial symmetry as the initial condition throughout the whole time interval $t\in[0,300]$, and besides its statistics are obviously different from those given by DNS.  This conclusion is clear and obvious from our previous publications \citep{Qin2024JOES, Liao2024NEC-arXiv, Liao2025JFM} about the 2D turbulent Kolmogorov flow.  However, as illustrated in this paper, when considering thermal fluctuation and/or stochastic environmental noise, both of CNS (with thermal fluctuation \& environmental disturbance) and DNS (that is badly polluted by numerical noise, but without thermal fluctuation \& environmental disturbance) give the same statistics, strongly suggesting that there should exist some relationships between numerical noise and thermal fluctuation \& environmental disturbance.  In this meaning, we highly suggest that the numerical noise should be approximately equivalent to thermal fluctuation or stochastic environmental disturbance, although further detailed investigations are certainly necessary in future.     

Note that a few molecular simulations using molecular-gas-dynamics (MGD) technique \citep{McMullen2022PRL} or  unified stochastic particle (USP) \citep{ma2024} demonstrated that thermal fluctuations might significantly influence the small-scale statistics of turbulence, leading to a $k$ scaling in the dissipation range of two-dimensional turbulent energy spectrum and a $k^2$ scaling in three-dimensional case.   This phenomenon has been confirmed by Landau-Lifshitz-Navier-Stokes (LLNS)  equations \citep{Bandak2022PRE} and/or fluctuating hydrodynamics  \citep{Bell2022JFM} but not by NS equation.    Note that DNS results of NS or LLNS equations are badly polluted by artificial numerical noises, as illustrated by \citet{Qin2024JOES}.       
Although LLNS equations \citep{Landau1959} are not as widely used as NS equations, it should be very interesting in future to use CNS to solve LLNS equations so as to study influence of thermal fluctuation on turbulence with negligible numerical noise.

\backsection[Acknowledgements]{Thanks to the anonymous reviewers for their valuable suggestions and constructive comments.}

\backsection[Funding]{This work is supported by State Key Laboratory of Ocean Engineering, Shanghai 200240, China}

\backsection[Declaration of Interests]{The authors report no conflict of interest.}

\backsection[Data availability statement]{The data that support the findings of this study are available on request from the corresponding author.}

\backsection[Author ORCID]{Shijun Liao, https://orcid.org/0000-0002-2372-9502; Shijie Qin, https://orcid.org/0000-0002-0809-1766}

\appendix

\section{Some definitions and measures}    \label{Key_measures}

For the sake of simplicity, the definitions of some statistic operators are briefly described below.
The spatial average is defined by
\begin{align}
& \langle\,\,\rangle_A=\frac{1}{4\pi^2}\int^{2\pi}_0\int^{2\pi}_0 dxdy,       \label{average_A}
\end{align}
the temporal average is defined by
\begin{align}
& \langle\,\,\rangle_{t}=\frac{1}{T_2-T_1}\int^{T_2}_{T_1} dt,       \label{average_t}
\end{align}
and the spatiotemporal average is defined by
\begin{align}
& \langle\,\,\rangle=\frac{1}{4\pi^2 (T_2-T_1)}\int^{2\pi}_0\int^{2\pi}_0\int^{T_2}_{T_1} dxdydt,       \label{average_all}
\end{align}
respectively, where $T_1=100$ and $T_2=300$ are chosen in the main text for an interval of time corresponding to a relatively stable state of the turbulent flow.

For the turbulent two-dimensional Kolmogorov flow considered in this paper, vorticity is given by the stream function
\begin{align}
& \omega(x,y,t)=\nabla^{2}\psi(x,y,t).       \label{vorticity}
\end{align}
We also focus on the kinetic energy
\begin{align}
& E(x,y,t) = \frac{1}{2}[u^2(x,y,t)+v^2(x,y,t)],    \label{kinetic_energy}
\end{align}
enstrophy
\begin{align}
& \Omega(x,y,t) = \frac{1}{2}\,\omega^2,    \label{enstrophy}
\end{align}
the kinetic energy dissipation rate
\begin{align}
& D(x,y,t)=\frac{1}{2Re}\sum_{i,j=1,2}\big [ \partial_iu_j(x,y,t)+\partial_ju_i(x,y,t) \big ]^2,    \label{dissipation_rate}
\end{align}
and enstrophy dissipation rate
\begin{align}
& D_\Omega(x,y,t)=\frac{1}{Re}|\nabla\omega|^2,    \label{enstrophy_dissipation_rate}
\end{align}
where $u_1(x,y,t)=u(x,y,t)$, $u_2(x,y,t)=v(x,y,t)$, $\partial_1=\partial /\partial x$, and $\partial_2=\partial /\partial y$.

The stream function can be expanded as the Fourier series
\begin{align}
& \psi(x,y,t)\approx\sum^{\lfloor N/3 \rfloor}_{\,m=-\lfloor N/3 \rfloor}\sum^{\lfloor N/3 \rfloor}_{\,n=-\lfloor N/3 \rfloor}\Psi_{m,n}(t) \exp(\mathbf{i}\,mx)\exp(\mathbf{i}\,ny),       \label{Fourier}
\end{align}
where $m$, $n$ are integers, $\lfloor\,\,\rfloor$ stands for a floor function, $\mathbf{i}=\sqrt{-1}$ denotes the imaginary unit, and for dealiasing $\Psi_{m,n}=0$ is imposed for wavenumbers outside the above domain $\sum$. Note that for the real number $\psi$, $\Psi_{-m,-n}=\Psi^*_{m,n}$ must be satisfied, where $\Psi^*_{m,n}$ is the conjugate of $\Psi_{m,n}$.
Therefore, the kinetic energy spectrum is defined as
\begin{align}
& E_k(t)=\sum_{k-1/2 \leq \sqrt{m^2+n^2} < k+1/2}\frac{1}{2}\,(m^2+n^2)\mid \Psi_{m,n}(t) \mid^2,       \label{kinetic_energy_spectrum}
\end{align}
where the wave number $k$ is a non-negative integer.
Noth that, if the stream function $\psi$ is obtained via the difference between two velocity fields, such as $\Delta \mathbf{u}=\mathbf{u}_{\mathrm{CNS}^*}-\mathbf{u}_{\mathrm{DNS}}$, the corresponding kinetic energy spectrum is denoted by $E_\Delta(k,t)$.

Filter-Space-Technique (FST) is employed in this investigation to extract the scale-to-scale energy and enstrophy fluxes, denoted as $\Pi_E^{[l]}$ and $\Pi_Z^{[l]}$ (see definitions below), respectively. FST, initially developed for large eddy simulation in the 1970s \cite{Leonard1975AG}, involves applying a low-pass filter to the velocity field. Mathematically, it is expressed as:
\begin{equation}
f^{[l]}(\mathbf{x},t) = \int G^{[l]}(\mathbf{x}-\mathbf{x}') f(\mathbf{x}',t) d \mathbf{x}',
\end{equation}
where $f$ represents $u$ or $v$ for the two-dimensional velocity field, $\mathbf{x}=(x, y)$ denotes the coordinate vector, and $G^{[l]}$ is chosen to be a round Gaussian filter for the scale $l$ \cite{Chen2003PRL,Boffetta2012ARFM,zhou2016JFM}.
For the incompressible Navier-Stokes equations, scale-to-scale energy and enstrophy fluxes can be derived analytically as:
\begin{equation}
\Pi_E^{[l]}= -\sum_{i,j=1,2} \left[ \left( u_iu_j \right)^{[l]} -u_i^{[l]}u_j^{[l]}\right]\frac{\partial u_i^{[l]}}{\partial x_j},
\end{equation}
\begin{equation}
\Pi_Z^{[l]}= -\sum_{i=1,2} \left[ \left( u_i\omega \right)^{[l]} -u_i^{[l]}\omega^{[l]}\right]\frac{\partial \omega^{[l]}}{\partial x_i},
\end{equation}
respectively.
Note that the sign of $\Pi_E^{[l]}$ or $\Pi_Z^{[l]}$ reveals the direction of energy or enstrophy transfer: a positive value indicates a cascade from the larger scale ($>l$) to the smaller scale ($<l$), i.e. the direct cascade, while a negative value signifies the reverse, i.e. the inverse cascade.



\bibliographystyle{jfm}
\bibliography{Kolmogorov2D}

\end{sloppypar}
\end{document}